\documentclass[twocolumn]{aastex63}
\usepackage{hyperref}
\usepackage{bm}
\usepackage{multirow}
\usepackage{gensymb}
\usepackage{enumitem}
\usepackage{tabularx}
\usepackage[flushleft]{threeparttable}
\usepackage[]{color}

\usepackage{natbib}
\usepackage{float}

\usepackage{amsfonts,amsmath,amssymb}
\usepackage{graphicx}
\usepackage{color}
\usepackage{xcolor}
\usepackage{natbib}
\usepackage{url}

\newcommand{\xstar}{\textsc{\scriptsize XSTAR}}

\newcommand{\athena}{\textsc{Athena\scriptsize ++ }}

\defcitealias{Dannen20}{D20}
\defcitealias{Waters21}{WPD21}
\defcitealias{Begelman83}{BMS83}
\def\CIVdbl{{\rm C~}\kern 0.1em{\sc iv}~$\lambda\lambda 1548, 1550$} 
\def\OVIIIi{\hbox{{\rm O}\kern 0.1em{\sc viii}}}
\def\SiXIVi{\hbox{{\rm Si}\kern 0.1em{\sc viii}}}
\def\OVIII{\hbox{{\rm O}\kern 0.1em{\sc viii}~{\rm Ly}$\alpha$}}
\def\SiXIV{\hbox{{\rm Si}\kern 0.1em{\sc viii}~{\rm Ly}$\beta$}}
\def\FeXXV{\hbox{{\rm Fe}\kern 0.1em{\sc xxv}}}
\def\FeXXVI{\hbox{{\rm Fe}\kern 0.1em{\sc xxvi}}}
\def\FeXXVK{\hbox{{\rm Fe}\kern 0.1em{\sc xxv}~{\rm K}$\alpha$}}
\def\FeXXVIK{\hbox{{\rm Fe}\kern 0.1em{\sc xxvi}~{\rm K}$\alpha$}}

\newcommand{\beq}{\begin{equation}}
\newcommand{\seq}{\end{equation}}

\renewcommand{\v}[1]{\ensuremath{\mathbf{#1}}} 
\newcommand{\gv}[1]{\ensuremath{\mbox{\boldmath$ #1 $}}} 
\renewcommand{\d}[2]{\frac{d #1}{d #2}} 
\newcommand{\pd}[2]{\frac{\partial #1}{\partial #2}}
\newcommand{\pdtext}[2]{\partial #1/\partial #2}
\newcommand{\f}{\frac}  
\newcommand{\e}[1]{\exp\left( #1 \right)} 
\newcommand{\bigb}[1]{\left[ #1 \right]} 
\newcommand{\bigp}[1]{\left( #1 \right)} 

\usepackage{letltxmacro,amsmath}
\LetLtxMacro{\originaleqref}{\eqref}
\renewcommand{\eqref}{Eq.~\originaleqref}

\submitjournal{ApJ}
\shorttitle{Dynamical TI in UFOs}
\shortauthors{Waters, Proga, Dannen, \& Dyda}
\begin{document}
\title{Dynamical thermal instability in highly supersonic outflows}

\correspondingauthor{Tim Waters}
\email{tim.waters@unlv.edu}
\author[0000-0002-5205-9472]{Tim Waters}
\affiliation{Department of Physics \& Astronomy \\
University of Nevada, Las Vegas \\
4505 S. Maryland Pkwy \\
Las Vegas, NV, 89154-4002, USA}
\affiliation{Theoretical Division, Los Alamos National Laboratory}
\author[0000-0002-6336-5125]{Daniel Proga}
\affiliation{Department of Physics \& Astronomy \\
University of Nevada, Las Vegas \\
4505 S. Maryland Pkwy \\
Las Vegas, NV, 89154-4002, USA}
\author[0000-0002-5160-8716]{Randall Dannen}
\affiliation{Department of Physics \& Astronomy \\
University of Nevada, Las Vegas \\
4505 S. Maryland Pkwy \\
Las Vegas, NV, 89154-4002, USA}

\author[0000-0002-1954-8864]{Sergei Dyda}
\affiliation{Department of Astronomy \\
University of Virginia \\
530 McCormick Rd \\
Charlottesville, VA 22904, USA}

\begin{abstract}
Acceleration can change the ionization of X-ray irradiated gas to the point that the gas becomes thermally unstable.  Cloud formation, the expected outcome of thermal instability (TI), will be suppressed in a dynamic flow, however, due to the stretching of fluid elements that accompanies acceleration.  It is therefore unlikely that cloud formation occurs during the launching phase of a supersonic outflow. In this paper, we show that the most favorable conditions for dynamical TI in highly supersonic outflows are found at radii beyond the acceleration zone, where the growth rate of entropy modes is set by the linear theory rate for a static plasma.   This finding implies that even mildly relativistic outflows can become clumpy, and we explicitly demonstrate this using hydrodynamical simulations of ultrafast outflows. We describe how the continuity and heat equations can be used to appreciate another impediment (beside mode disruption due to the stretching) to making an outflow clumpy: background flow conditions may not allow the plasma to enter a TI zone in the first place. The continuity equation reveals that both impediments are in fact tightly coupled, yet one is easy to overcome.  Namely, time variability in the radiation field is found to be a robust means of placing gas in a TI zone.  We further show how the ratio of the dynamical and thermal timescales enters linear theory; the heat equation reveals how this ratio depends on the two processes that tend to remove gas from a TI zone --- adiabatic cooling and heat advection.
\end{abstract}

\section{Introduction}
High-velocity, supersonic clouds are commonly invoked to explain various
spectral properties of active galactic nuclei 
\citep[AGN; for an overview, see][]{Krolik99, Netzer13}. 
Such clouds are envisioned to coexist with more highly ionized plasma. 
However, the strong radiation field, the presence 
of velocity shear, and cloud collisions, among other processes, 
make clouds relatively short lived 
entities \citep[][]{Krolik77, Krolik79,Krolik88, Mathews77,Mathews86}.
This conclusion from early studies of AGN clouds was later verified
by hydrodynamical investigations that employed a local approximation 
\citep{Proga14, PW15}.  In followup investigations also employing local simulations \citep{WP16,Waters17}, we nevertheless concluded that cloud disruption itself can serve as an efficient regeneration mechanism because turbulent fluctuations generated by the flow supply new perturbations 
that can condensate due to classical thermal instability \citep[TI;][]{Field65}.

As first illustrated by \citet{Field65}, however,
the linear theory of classical TI becomes complicated in inhomogeneous flows.  A local approximation no longer suffices to understand the linear growth, let alone the nonlinear behavior.
The much richer theory of `dynamical TI' has been investigated analytically in many subsequent papers
\citep[e.g.,][]{Mathews78, Balbus86, Nulsen86, Krolik88, Balbus89, Mathews90,Balbus16}.
In recent studies employing global rather than local simulations, we sought to understand why outflows tend to avoid becoming multiphase due to dynamical TI \citep[see][hereafter, D20 and WPD21, respectively]{Dannen20,Waters21}. 
There are two main reasons that can be understood analytically (see \S{2}) 
and stated in physical terms: 
(i) a given fluid element could start off in a region where it is thermally unstable but before undergoing exponential growth to nonlinear amplitudes, it could be advected into a stable region where growth can be `undone' by exponential decay;
(ii) the linear perturbations most susceptible to TI (so-called entropy modes) could be disrupted due to the stretching of fluid elements undergoing acceleration.
We showed that in the case of thermally driven winds, favorable 
conditions for dynamical TI only occur at parsec scales in AGN.

These studies of multiphase gas dynamics in AGN take on new relevance
with a growing number of observations establishing that ultrafast outflows (UFOs),
systems with high-ionization blueshifted absorption lines in the range $0.03-0.3\,c$,
are rather commonplace \citep{Tombesi13,Laha20}.
While the early identification of UFOs were mainly in the hard X-ray band, and only in quasars \citep{Chartas02,Pounds03,Reeves03}, there had been at least one report of mildy relativistic soft X-ray absorption in a less luminous AGN \citep{Leighly97}.  Evidence for both soft and hard UFO components in Seyfert galaxies has accumulated only in the last decade, following initial studies using observations from \textit{Chandra} \citep{Gupta13,Gupta15} and \textit{XMM-Newton} \citep{Longinotti15}.
Upon identifying mildy relativistic soft X-ray absorption in a quasar also --- in the nearby system PDS 456  $(z = 0.184)$, where conclusive evidence for a UFO was previously only from Fe K-shell lines --- \citet{Reeves09} and \citet{Matzeu16} suggested that this UFO might be clumpy.  This suggestion was also made by \citet{Hamann18}, who identified a UV absorber at $\sim 0.3\,c$ in this system.  
Remarkably, follow up observations have placed the soft X-ray absorber and the iron K absorber at the same velocity \citep{Reeves20}, strengthening a similar finding made already for PG 1211+143 \citep{Reeves18}.  

These observations make dynamical TI the most plausible mechanism to
account for the clumpiness of UFOs because alternative explanations such as cloud entrainment \citep[e.g.,][]{Serafinelli19} or a cooling, shocked flow \citep[e.g.,][]{Pounds13} would necessarily require the lower ionization absorber to have significantly slower speeds \citep[e.g.,][]{Wagner13}.
Moreover, the soft X-ray lines that would trace the clumps are highly
variable in both PDS 456 \citep{Matzeu16, Reeves20} and PG 1211+143 \citep{Reeves18}. 
This is the expected outcome if,
consistent with the AGN cloud studies mentioned above, 
clumps are highly dynamical entities
that are nevertheless efficiently regenerated to maintain a low ionization component.
 
Our recent models of clumpy thermally driven winds
led us to identify a rule of thumb that should apply to self-consistent models of UFOs:
the acceleration required to produce highly supersonic AGN winds effectively stabilizes the flow to dynamical TI due to reasons (i) and (ii) above \citepalias{Dannen20,Waters21}.
On this basis, it would be tempting to conclude that UFOs should not be clumpy on sub-parsec scales and hence that the observed variability is due to either ionization changes caused by a variable accretion rate or perhaps by dynamics associated with shocks.  
However, we show in this paper that there is in fact a wide parameter space
conducive to clump formation in fast AGN outflows, namely all distances beyond the acceleration zone where the outflows can coast upon approaching a terminal velocity.     

This paper is organized as follows.
In \S{2}, 
we extend the theory of dynamical TI to highly supersonic outflows. 
In \S{3},
we present illustrative hydrodynamical simulations of UFOs to show that in regions of near constant velocity, irradiated flows can become clumpy provided they can enter and spend sufficient time within a TI zone.    
In \S{4}, we discuss our results in the context of our previous work, 
before noting their observational implications.  
We conclude in \S{5} with a summary of our results and a unifying principle to explain why Compton heated winds and UFOs are both prone to becoming clumpy due to dynamical TI. 

\section{Dynamical TI}
\label{sec:theory}
Photoionization modeling studies of AGN typically assume that gas will reach thermal equilibrium, i.e. occupy points on the S-curve (defined as the contour $\mathcal{L} = 0$ with $\mathcal{L}$ being the net sum of all radiative heating and cooling processes) in the phase space spanned by the temperature and the ionization parameter $\xi = 4\pi F_X/n_H$, where $F_X$ is the flux of ionizing radiation and $n_H$ is the number density of hydrogen.  In fast flows, this is unlikely to be the case due to adiabatic cooling \citep[see][]{Kallman10}.  
When the flow deviates from the S-curve, the criterion for thermal instability is that found by \citet{Balbus86} rather than by \citet{Field65}. 

These \textit{local} instability criteria both follow from the globally applicable amplification solution for radial flows first derived by \citet{Mathews78} and further generalized by \citet[][see also \citealt{Krolik88}]{Balbus89},  
 \begin{equation} \label{eq:growth}
 \f{\delta \rho}{\rho}= a_0 \exp\bigp{- \int^t_{t_0} \f{N_p}{\gamma t_{\rm cool}}\, dt'}.
 \end{equation}
Here, $a_0$ is the initial perturbation amplitude, 
$t_{\rm cool} = \mathcal{E}/\Lambda_0$ is a characteristic cooling time with 
$\mathcal{E} = c_{\rm v} T$ the internal energy per unit mass 
of the gas\footnote{In this work, we assume an adiabatic index $\gamma = 5/3$ and mean particle mass $\bar{m} = \mu m_p$ with $\mu = 0.6$, so the specific heat at constant volume is $c_{\rm v} = (\gamma - 1)^{-1}k/\bar{m} = 2.0636\times10^8~{\rm cm^2\, s^{-2}\,K^{-1}}$.}, and $\Lambda_0$ is a characteristic cooling rate (in units of ${\rm erg~s^{-1}\,g^{-1}}$) that is used to define the dimensionless growth rate
\beq \label{eq:Np}
N_p \equiv \f{T^2}{\Lambda_0}\bigb{\f{\partial (\mathcal{L}/T)}{\partial T}}_p,
\seq 
where the subscript $p$ denotes evaluation of a quantity at constant pressure.
Because $N_p/(\gamma\, t_{\rm cool})$ is the solution to the cubic dispersion relation for local TI in the isobaric limit \citep[see][whose notation we adopt here]{Waters19a},
it was pointed out by \citet{MP13} that \eqref{eq:growth} is a very intuitive result: the growth rate of a co-moving radial perturbation is found by tracking the instantaneous local isobaric growth rate encountered by the perturbation as it propagates for a duration $\Delta t = t - t_0$.
When the background flow is steady and homogeneous, \eqref{eq:growth} reduces to the 
local isobaric amplification solution, $\delta \rho/\rho = a_0\,\exp[-(N_p/\gamma) \Delta t/t_{\rm cool}]$.  Instability results if $N_p < 0$, so from \eqref{eq:Np} we recover the
local instability criterion of \citet{Balbus86}, $[\pdtext{(\mathcal{L}/T)}{T}]_p < 0$, 
applying to a fluid element that is either in ($\mathcal{L} = 0$) 
or out ($\mathcal{L} \neq 0$) of thermal equilibrium, the former case reducing to that of \citet{Field65}, $[\pdtext{\mathcal{L}}{T}]_p < 0$.  

\eqref{eq:growth} quantifies the flow effects that we mentioned in \S{1}. 
For example, during the time evolution a moving perturbation could be 
in a region where $N_p < 0$, what we hereafter refer to as a TI \textit{zone}.  
The density contrast will grow exponentially 
at the rate $|N_p|/(\gamma\, t_{\rm cool})$, provided the entropy mode 
is not disrupted by a deformation of the fluid element carrying it.  
At some later time, the perturbation could enter a region with $N_p > 0$,
meaning the gas is stable and the perturbation decays exponentially 
(at the rate $|N_p|/(\gamma\, t_{\rm cool})$ evaluated at this new location).  
If passage through a TI zone results in the formation of a cloud 
(i.e. the saturation of TI), the cloud itself then serves as a new background flow 
in which perturbations can separately grow or damp depending 
on the parameter space occupied by this multiphase gas.

Mode disruption is most easily understood by examining the 
continuity equation in the form $D\ln\rho/Dt = -\nabla \cdot \gv{v}$.
Stretching terms, e.g. $\pdtext{v_z}{z}$ for vertical, planar flow and  
$\pdtext{v_r}{r}$ for radial flow, act to reduce the density --- 
literally to stretch the wavelength in the case of a perturbation 
or to expand a clump as it is advected along.  
By focusing our attention in this work on the supersonic part 
of the  outflow that is fully accelerated, these stabilizing dynamical terms
are absent.  Specifically, we explore cases in which the scale length 
for changes in velocity,
\beq \lambda_{v} \equiv (d\ln v/dl)^{-1} , \label{eq:lambda_v} \seq
is very large, i.e., $\lambda_{v} \gg r_0$, where $v$ 
is the flow velocity, $r_0$ is the distance to the base of a streamline, 
and (to not confine our results to radial flows) we consider flow properties 
along a streamline parameterized by $l$, the distance from $r_0$.

Using $dl = v\,dt$, we can express \eqref{eq:growth} in a form more suitable 
for analysis:
 \beq \label{eq:growth2}
 \f{\delta \rho}{\rho}= a_0 \exp\bigp{- \f{1}{r_0} \int^l_{0} \sigma_p(l')\,\f{t_{\rm dyn}(l')}{t_{\rm th}(l')}\, dl'}.
 \seq
Here, $t_{\rm dyn} = r_0/v$ is the dynamical timescale 
and we have introduced 
$\sigma_p(l) \equiv (\Lambda_0/|\mathcal{L}|) N_p/\gamma$ to remove
the density $\rho = \bar{m}\,n$ from within the dimensionless growth rate 
(through $\mathcal{L} = (n^2/\rho) L_{\rm net} =  (n/\bar{m}) L_{\rm net}$, with $L_{\rm net}$ 
being the net cooling rate in units of ${\rm erg~cm^3\,s^{-1}}$).  
Expanding the derivative in \eqref{eq:Np} explicitly gives
\beq 
\sigma_p(l) = \f{1}{\gamma}\f{T}{|L_{\rm net}|}\left[\left(\pd{L_{\rm net}}{ T}\right)_p - 2\f{L_{\rm net}}{T}\right]. 
\label{eq:sigma_p}
\seq  
With this expression for the growth rate, the density dependence appears
in the thermal timescale,
\beq
t_{\rm th} \equiv  \f{\mathcal{E}}{|\mathcal{L}|}  = \f{\Lambda_0}{|\mathcal{L}|} t_{\rm cool}.
\label{eq:t_th}
\seq

This formulation is advantageous not only because we directly work with tabulated values of $L_{\rm net} = L_{\rm net}(\xi,T)$ but also to make a relevant connection with the broader literature on dynamical TI simulations, where the ratio $t_{\rm dyn}/t_{\rm th}$ has been identified as the main determinant of whether TI results in multiphase gas production \citep[e.g.,][]{Sharma12, McCourt12, Gaspari12}. 
However, it is clear from \eqref{eq:growth2} that it is the product $\sigma_p(l)\, t_{\rm dyn}/t_{\rm th}$, \textit{rather than the ratio $t_{\rm dyn}/t_{\rm th}$ itself}, that determines whether or not TI will lead to clump formation.
To elucidate the role of $t_{\rm dyn}/t_{\rm th}$ alone, we can place a theoretical bound on this ratio by first defining the
effective distance into the TI zone where the maximum value of $|\sigma_p(l)|$ is encountered.  Denoting this maximum growth rate by $|\sigma_p|_{\rm max}$, this distance is
\beq 
{\Delta l}_{\rm TI, max}(l) = \f{1}{|\sigma_p|_{\rm max}} \int_{l_{\rm TI}}^l |\sigma_p(l')|\, dl',
\label{eq:Delta_l}
\seq 
where $l_{\rm TI}$ denotes the start of the TI zone (the location where $\sigma_p(l)$ first becomes negative).
By \eqref{eq:growth2}, reaching $\delta \rho/\rho \gtrsim 1$ requires the maximum value of $t_{\rm dyn}/t_{\rm th}$ to exceed $|\sigma_p|_{\rm max}^{-1} \ln(1/a_0)\, r_0/{\Delta l}_{\rm TI, max}$, which will be guaranteed if the minimum value exceeds it also.
Linear stability by definition entails having perturbation amplitudes $a_0 \lesssim 0.1$, hence enforcing $\ln(1/a_0) \gtrsim \ln(10) \approx 2.3$ gives
\beq 
\f{t_{\rm dyn}}{t_{\rm th}} \gtrsim \f{2.3}{({\Delta l}_{\rm TI, max}/r_0)|\sigma_p|_{\rm max} }.
\label{eq:tscale_bound}
\seq
This bound is consistent with the empirical result obtained in the above cited studies of dynamical TI in stratified atmospheres when $({\Delta l}_{\rm TI, max}/r_0)|\sigma_p|_{\rm max} \sim 1$.  It furthermore shows how, depending on the cooling function and background flow dynamics, the production of multiphase gas may require $t_{\rm dyn}/t_{\rm th} \gg 1$ or permit $t_{\rm dyn}/t_{\rm th} \ll 1$, which might explain the recent finding by \citet{Esmerian21} that this ratio is a poor predictor of when dynamical TI leads to multiphase structure in simulations of galaxy halos.  

Notice in particular that it can be necessary to have $t_{\rm th} \ll t_{\rm dyn}$ for two different reasons: 
(i) the growth rate distribution an entropy mode encounters along its path can be small overall with $\sigma_p(l) \ll 1$ and yet sharply peaked about $|\sigma_p|_{\rm max} \sim 1$, to give ${\Delta l}_{\rm TI, max} \ll r_0$; 
or (ii) even the maximum growth rate is very small, having $|\sigma_p|_{\rm max} \ll 1$. On the other hand, for dynamical TI to operate when $t_{\rm th} \gg t_{\rm dyn}$ requires either ${\Delta l}_{\rm TI, max} \gg r_0$ (for $|\sigma_p|_{\rm max} \sim 1$) or a net cooling function that gives $|\sigma_p|_{\rm max} \gg 2.3 ({\Delta l}_{\rm TI, max}/r_0)^{-1}$.
As will be shown in \S{3.3}, our UFO model is in this last category with $|\sigma_p|_{\rm max} \gg 1$.

Plugging in characteristic values for UFOs, we find
\beq
\f{t_{\rm dyn}}{t_{\rm th}} = 2.38 \, M_7 \left(\f{r_0/r_g}{10^4}\right)
\left|\f{L_{\rm net}}{10^{-24}}\right|
n_7\, v_{0.1}^{-1}\,T_5^{-1},
\label{eq:trat}
\seq 
where $M_7 = M_{\rm bh}/10^7\,M_\odot$, $n_7 = n/10^7 {\rm cm}^{-3}$, $v_{0.1} = v/0.1\,c$, $T_5 = T/10^5\,{\rm K}$, and $r_g = GM_{\rm bh}/c^2$.
Comparing with \eqref{eq:tscale_bound}, we conclude that there is sufficient time for entropy modes to be exponentially amplified provided the TI zone has (${\Delta l}_{\rm TI}/r_0) |\sigma_p|_{\rm max} \gtrsim 1$.
However, regardless of how large or small $t_{\rm dyn}/t_{\rm th}$ is, gas must occupy a TI zone to be thermally unstable, a requirement that is by no means guaranteed in an outflow.

\begin{figure*}
  \centering
  \includegraphics[width=0.7\textwidth]{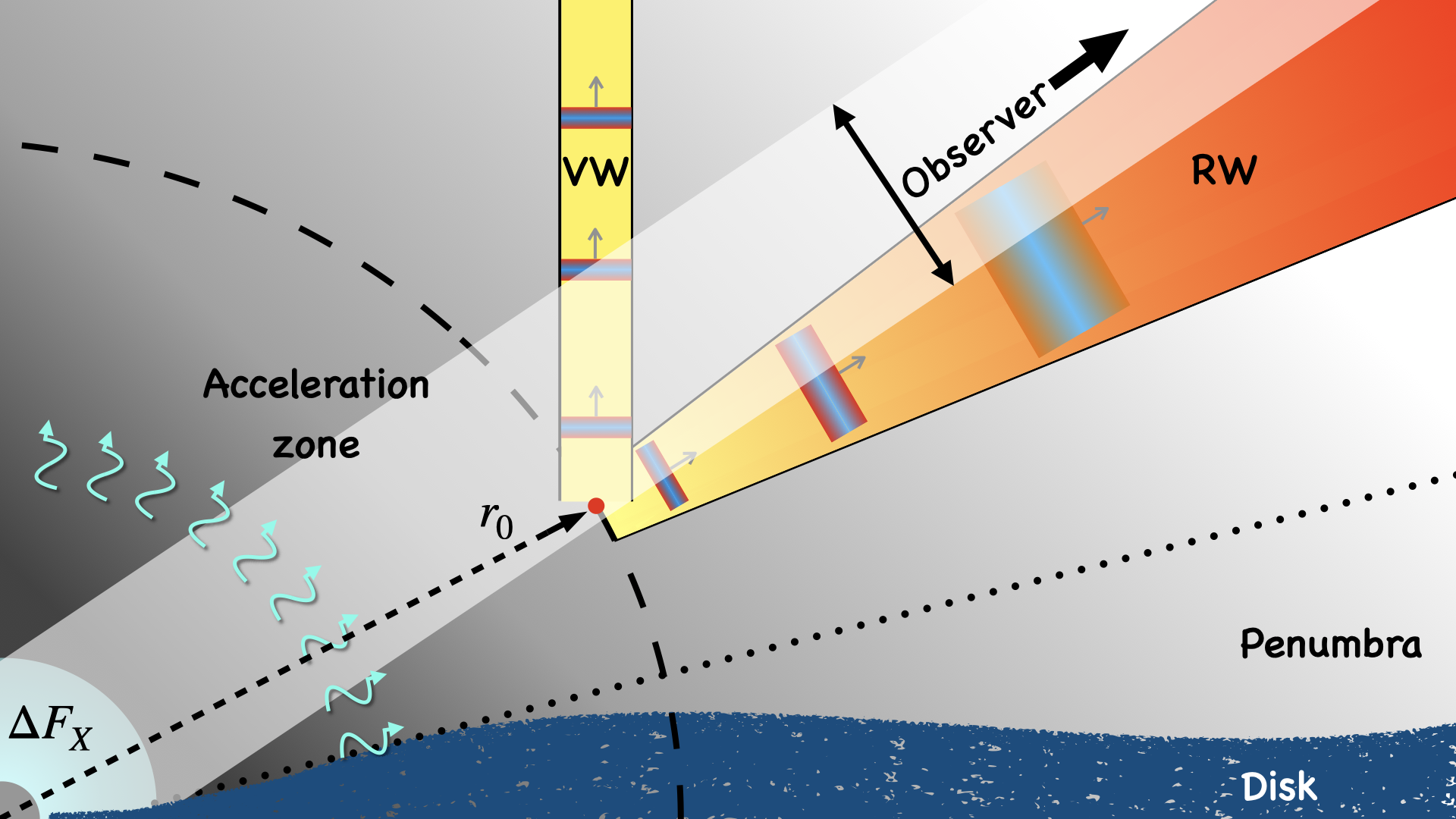}
  \caption{
Sketch of the global environment of the UFO model explored here.
The gray background represents a radiation field with flux varying as $r^{-2}$.  The physical scenario that we model is a pre-existing highly supersonic outflow settling into a new steady state after an increase in the ionizing flux (illustrated by the cyan region and wiggly arrows).  We demonstrate that this rise in flux can cause the plasma to enter a TI zone.  This is one of three necessary conditions for dynamical TI to robustly operate. The other two are 
(i) the ratio $t_{\rm dyn}/t_{\rm th}$ must be sufficiently large --- see \eqref{eq:tscale_bound} --- for the outflow to occupy the TI zone long enough for perturbations to become nonlinear; and 
(ii) the TI zone must reside beyond the outflow’s acceleration zone (here shown by the dashed circle at $r=r_0$) so that entropy modes are not disrupted by stretching terms.  
We obtain clumpy outflow solutions using a numerical setup for two flow tubes with different cross-sectional areas. For both geometries, we solve for the dynamics along streamlines originating from the same base location (marked by a red dot) with the same velocity.  
`VW’ streamlines are vertically oriented and occupy a constant area, giving $\rho = const$ by \eqref{eq:rhoAv}.
`RW’ streamlines expand as $A\propto r^2$ so that $\rho \propto r^{-2}$.  The qualitative effect these background density profiles have on newly formed clouds is drawn.  To show that clouds are expected to produce variable absorption lines, an observer’s view of the X-ray source is depicted (the arrow indicates the diameter of the unresolved background continuum emission region).  The location of a shadowed region (the penumbra) is included to indicate that the rise in flux can be caused by `deshadowing' rather than by an increase in the luminosity. 
  }
  \label{fig:sketch}
\end{figure*}

\subsection{Entering and occupying TI zones}
Both the heat equation and the continuity equation can be used to appreciate the other impediment (i.e. beside mode disruption due to the stretching of fluid elements) to making an outflow clumpy: 
background flow conditions may not allow the plasma to enter a TI zone in the first place.
As shown below, the continuity equation reveals that both impediments are in fact tightly coupled, while the heat equation
can be manipulated to show how the ratio $t_{\rm dyn}/t_{\rm th}$ depends on the two obstacles encountered: adiabatic cooling and heat advection.  To understand the effects of the latter processes, we isolate each one by considering the two separate flowtube geometries shown in Fig.~\ref{fig:sketch}.  
This sketch depicts the global environment consistent with the UFO scenario we model, namely a supersonic wind that has approached its terminal velocity and reached a steady state.  The continuity equation is therefore 
\begin{equation} \label{eq:rhoAv}
\rho\, v A= \dot{M},
\end{equation} 
where $A$ is the cross-sectional area of the flowtube and $\dot{M}=constant$ is the mass loss rate. The steady state heat equation reads
\beq 
\gv{v}\cdot \nabla \mathcal{E} = -(\gamma-1) \mathcal{E} \nabla \cdot \gv{v} -  \mathcal{L}.
\label{eq:heat}
\seq
Notice that the vertical wind (VW) flowtube has $A = constant$, and thus $\rho = constant$ by \eqref{eq:rhoAv} once $v \approx constant$. It then follows that $\nabla \cdot \gv{v} = 0$ in \eqref{eq:heat}.  In the absence of adiabatic cooling, only the advection term $\gv{v}\cdot \nabla \mathcal{E}$ can balance $\mathcal{L}$ in the VW flowtube.
The radial wind (RW) flowtube, meanwhile, has $A \propto r^{2}$, so this outflow has $\rho \propto r^{-2}$ once $v \approx constant$.  As will be made clear below, in this case the expansion of the flowtube makes the adiabatic cooling term dominant over the advection term.

\begin{figure*}
  \centering
  \includegraphics[width=\textwidth]{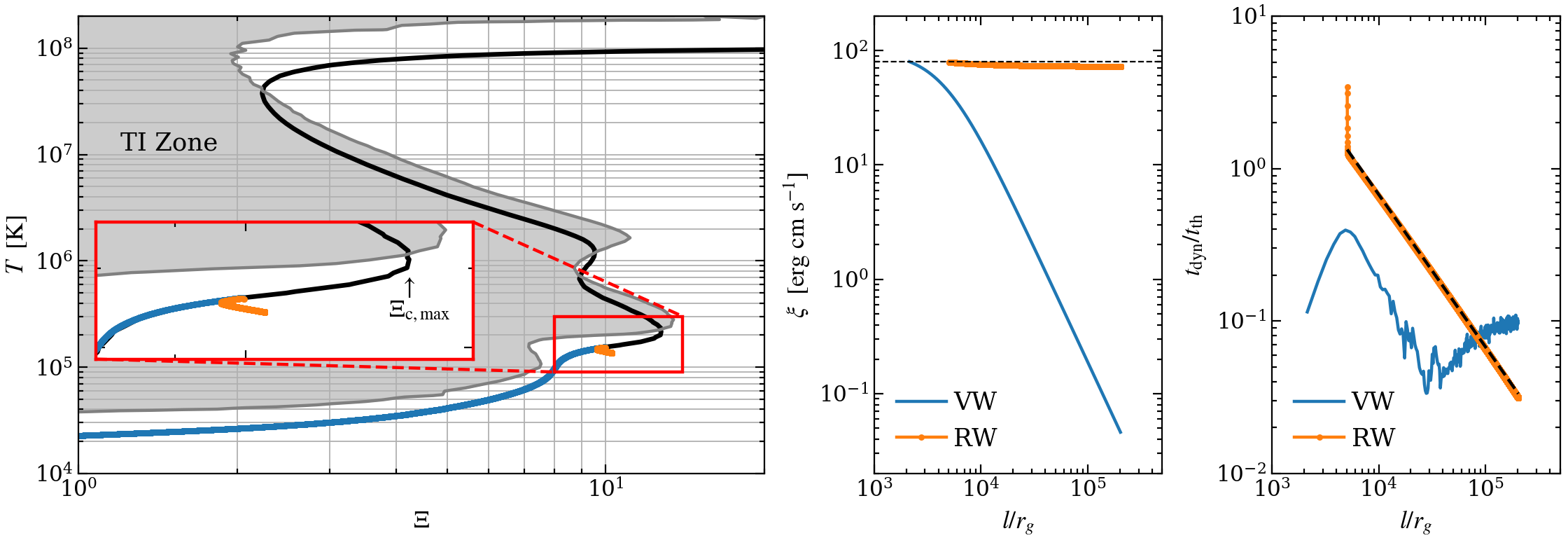}
  \caption{
  Steady state solutions for the RW flowtube (orange curves) and the VW flowtube (blue curves) of our illustrative UFO model in Fig.~\ref{fig:sketch}.  
  \textit{Left panel}: solution tracks on the phase diagram.  The S-curve is shown in black, and the TI zone (gray region) is where $N_p < 0$; its boundary is the `Balbus contour' where $N_p =0$.  Both solutions undergo cooling and hence do not enter the TI zone.  The inset panel shows that the RW solution falls below the S-curve.   
  \textit{Middle panel}: ionization parameter versus distance along the streamline.  Since $v \approx constant$ and $F_{\rm X} \propto r^{-2}$ in both outflow geometries, the RW solution has $\xi \approx constant$  by \eqref{eq:FvA} because $A\propto r^2$, while the VW solution has $\xi \approx r^{-2}$ because $A=constant$.  
  \textit{Right panel}: $t_{\rm dyn}/t_{\rm th}$ versus distance along the streamline.  In the RW solution, the cooling that places the solution below the S-curve is almost entirely due to adiabatic expansion; the black dashed line shows the adiabatic cooling relation $t_{\rm dyn}/t_{\rm th} = (\gamma-1) r_0/|\lambda_\rho|$ (see \eqref{eq:heat3}).  In the VW solution, cooling is solely due to heat advection, hence the VW curve is the quantity $r_0/|\lambda_T|$.
  }
  \label{fig:A0}
\end{figure*}

In Fig.~\ref{fig:A0}, we plot the solutions to \eqref{eq:rhoAv} and \eqref{eq:heat} that are obtained in \S{3} by solving the full time-dependent equations of non-adiabatic gas dynamics for both the VW and RW flowtubes.  In the left panel, these solutions are plotted in phase space on the $(T,\Xi)$-plane, where $\Xi \equiv (F_X/c)/p$ is the `pressure' ionization parameter (the ratio of the radiation pressure due to $F_X$ and the gas pressure).  The black contour is the S-curve (where $L_{\rm net} = 0$) and the shaded gray region\ is the TI zone (defined by $\sigma_p < 0$; see \eqref{eq:sigma_p}).  
The boundary of the TI zone (where $\sigma_p = 0$) defines the Balbus contour \citep[see][]{PW15}, and it intersects the S-curve at points where the slope of the S-curve changes sign.  While the slope of the S-curve and the width of the gap between this curve and the Balbus contour depend on the net cooling rates, which in turn are derived for gas of a given metallicity ionized by a particular radiation field, the main conclusions of this paper are expected to hold so long as the TI zone extends to the right of the S-curve.  The net cooling function $L_{\rm net}$ was calculated using \xstar~\citep{Bautista01,Kallman01} assuming solar metallicity  from the unobscured spectral energy distribution (SED) of NGC 5548 obtained by \citet{Mehdipour15} (for details, see \citealt{Dyda17} and \citealt{Dannen19}).  

\eqref{eq:rhoAv} and \eqref{eq:heat} can be further examined to understand why plasma in either flowtube cannot enter a TI zone if initially located on the cold stable branch of an S-curve (locations having $\Xi < \Xi_{\rm c,max}$).  Entering a TI zone requires climbing the S-curve, i.e. for $\Xi$ (and therefore $\xi$) to be an increasing function of $r$.  Considering that $\xi \propto F_{\rm X}/\rho$, we see from \eqref{eq:rhoAv} that streamlines in steady state regions of a flow have 
\beq \xi \propto F_X\,v\,A .\label{eq:FvA}\seq 
For an optically thin gas, $F_{\rm X}$ usually decreases with $l$; in the most commonly explored case of a point X-ray source with constant luminosity, $F_{\rm X} \propto r^{-2}$.
Thus, unless $v$ is \textit{not constant}, irradiation from a point source results in the scaling
$\xi \propto r^{-2}$ for VW streamlines and $\xi = constant$ for RW streamlines.  These scalings approximate very well the self-consistently obtained profiles that are shown in the middle panel of Fig.~\ref{fig:A0}.\footnote{Note that because the flow speed is not perfectly constant (it decelerates slightly), $\xi$ is decreasing somewhat in the RW case.} 

Herein lies the tight coupling between the two impediments to obtaining clumpy outflow solutions: entering a TI zone generally requires that an outflow be accelerating (i.e. $\xi$ will increase due to  $\xi \propto v$ in \eqref{eq:FvA}), yet acceleration has a stabilizing effect on TI by preventing exponential growth.  We can make this point differently by differentiating \eqref{eq:FvA} with respect to $l$.  Using \eqref{eq:lambda_v},
\begin{equation} \label{eq:xi_cond}
 \d {\ln \xi}{l}= \d {\ln F_{\rm X}}{l} + \d{\ln A}{l} + \f{1}{\lambda_{v}}.
\end{equation} 
This relation states that significant increases in $F_{\rm X}$, A, or $v$ are needed to have $d\xi/dl > 0$. Not only are small values of $\lambda_{v}$ disruptive to entropy modes, but they keep $d\xi/dl >0$ upon entering the TI zone, making it difficult for gas to remain there for very long.

We can next understand why the VW solution occupies the S-curve while the RW solution deviates below it into a region of net heating (see the `tracks' of blue and orange points on the inset panel of Fig.~\ref{fig:A0}).  For gas in the VW flowtube to stay close to thermal equilibrium by following the S-curve and yet satisfy $\xi \propto r^{-2}$, it is necessary for heat to be efficiently transported downstream.  On the other hand, because $\xi$ is nearly constant in the RW solution, the temperature decrease is small and very little heat is being advected.  Intuitively, we expect the flow to fall below the S-curve due to strong adiabatic cooling only when $t_{\rm dyn} < t_{\rm th}$ --- is this the case?  \eqref{eq:heat} can be written as a useful relation for $t_{\rm dyn}/t_{\rm th}$ by dividing by $\mathcal{E}$ and eliminating $\nabla \cdot \gv{v}$ using the continuity equation (in the form $\nabla \cdot \gv{v} = - \gv{v} \cdot \nabla \ln \rho$):
\beq
\gv{v}\cdot \nabla \ln\mathcal{E} = (\gamma-1) \gv{v}\cdot \nabla \ln \rho -  \f{\mathcal{L}}{\mathcal{E}}.
\label{eq:heat2}
\seq 
Upon introducing the temperature and density scale lengths $\lambda_T \equiv (\nabla \ln \mathcal{E})^{-1}$ and $\lambda_\rho \equiv (\nabla \ln \rho)^{-1}$,  \eqref{eq:heat2} can be rearranged to give the following identity for the flow along streamlines:
\beq
\f{t_{\rm dyn}}{t_{\rm th}} = \left|(\gamma-1)\f{r_0}{\lambda_\rho} -  \f{r_0}{\lambda_T} \right|.
\label{eq:heat3}
\seq 
This dimensionless form of the heat equation again shows why it is useful to consider the separate flowtube geometries in Fig.~\ref{fig:sketch}: VW streamlines have $\lambda_\rho = \infty$, hence $t_{\rm dyn}/t_{\rm th} = r_0/|\lambda_T|$, 
while RW streamlines have $|\lambda_\rho| \ll |\lambda_T|$ giving $t_{\rm dyn}/t_{\rm th} \approx (\gamma-1) r_0/|\lambda_\rho|$.

\eqref{eq:heat3} can now be used to gauge if the rate of adiabatic cooling (quantified by the ratio $r_0/\lambda_\rho$) is so high that the flow cannot remain on the S-curve.  In the absence of heat advection, this will evidently be the case when $|\lambda_\rho| \gtrsim r_0$, for then $t_{\rm dyn}/t_{\rm th} \lesssim 1$.
Indeed, the RW flowtube has $\lambda_\rho \approx -r/2$, 
and thus $t_{\rm dyn}/t_{\rm th} \approx 2 (\gamma-1) r_0/r$ by \eqref{eq:heat3}.
It is now a priori evident why `tracks' of the RW solution on the phase diagram must fall increasingly below the S-curve at larger $r$: in \eqref{eq:heat}, net radiative heating is needed to balance the strong adiabatic cooling.  The black dashed line in the right panel of Fig.~\ref{fig:A0} is the relation $2 (\gamma-1) r_0/r$, which nearly
equals $|t_{\rm dyn}/t_{\rm th}|$ in the RW case (orange curve) everywhere except for the innermost region, which on the phase diagram corresponds to the points along the S-curve. 

The behavior shown in Fig.~\ref{fig:A0} is clearly not conducive to placing gas in the TI zone. Therefore, one needs to consider additional physical processes that would allow gas to `climb' the S-curve.  

\begin{table*}
\centering
\begin{threeparttable}
\small
\caption{UFO model parameters for the background flow solutions}   
\begin{tabularx}{0.81\linewidth}{l c c c c} 
\toprule
\label{table}      
 & & Stable solutions\tnote{$\dagger$} & \multicolumn{2}{c}{Unstable solutions\tnote{$\ddagger$}} \\
Quantity & Symbol & Either flowtube & RW flowtube & VW flowtube \\
\hline
\underline{\textit{GOVERNING PARAMETERS}} \\
Eddington fraction (increment\tnote{*})   & $\Gamma_0$ ($\Delta\Gamma$) & $10^{-3}$ (0) & $10^{-3}$ ($2\times10^{-3}$) & $10^{-3}$ ($2\times10^{-3}$)  \\    
Pressure ionization parameter  & $\Xi_0$  & 9.95 & 10.43 & 9.56\\
Hydrodynamic escape parameter    & ${\rm HEP}_0$  & $5\times10^3$   & $1.73\times10^3$ & $1.60\times10^3$\\
Mach number & $\mathcal{M}_0$ & 250    & 148 & 142 \\
\hline
\underline{\textit{DERIVED PARAMETERS}}\tnote{**} \\
Radial distance to wind base & $r_0$  & $5.08\times10^3 r_g$ & $5.08\times10^3 r_g$ & $5.08\times10^3 r_g$\\
Number density               & $n_0$  & $1.93\times10^7\,\rm{cm^{-3}}$ & $1.90\times10^7\,\rm{cm^{-3}}$ & $1.93\times10^7\,\rm{cm^{-3}}$\\
Injection velocity & $v_0$ & $0.0495\,c$ & $0.0495\,c$ & $0.0495\,c$ \\
Density ionization parameter  & $\xi_0$ & $80\,{\rm erg\,cm\,s^{-1}}$ & $238\,{\rm erg\,cm\,s^{-1}}$ & $239\,{\rm erg\,cm\,s^{-1}}$ \\
Temperature         & $T_0$ & $1.55\times10^5\,{\rm K}$ &$4.38\times10^5\,{\rm K}$ &$4.80\times10^5\,{\rm K}$\\
Dynamical time        & $t_{\rm dyn}(r_0)$ & 306~days & 306~days & 306~days\\
Thermal time         & $t_{\rm th}(r_0)$ & \tnote{***} & 101~days & 4.89~years\\
Net cooling rate $[{\rm erg~cm^3\,s^{-1}}]$     & $L_{\rm net}(r_0)$ & \tnote{***} & $4.74\times10^{-25}$ & $3.34\times 10^{-26}$\\
\hline \hline
\end{tabularx}

\begin{tablenotes}
\item[$\dagger$] Values of flow quantities at $r_0$. 
\item[$\ddagger$] Values of flow quantities at $r_p = 1.01\,r_0$ (the location where a perturbation is applied) in the new steady state reached after raising the flux by increasing $\Gamma$.  These values are obtained numerically.
\item[*] Note: `increment' refers to the increase in $\Gamma$ used to place stable flow into the TI zone; see \eqref{eq:ramp}.
\item[**] To be consistent with the SED of \citet{Mehdipour15} that was used to derive $L_{\rm net}$, we adopt the black hole mass of NGC 5548, $M_{\rm bh} = 5.2\times 10^7\, M_\odot$.  This gives $r_g=GM_{\rm bh}/c^2 = 7.72\times10^{12}\,{\rm cm}$.  
\item[***] These values do depend on the flowtube; for RW, $t_{\rm th}(r_0) = 118~{\rm days}$ and $L_{\rm net}(r_0) = 1.61\times 10^{-25} {\rm erg~cm^3\,s^{-1}}$, while for VW, $t_{\rm th}(r_0) = 9.67~{\rm years}$ and $L_{\rm net}(r_0) = -5.43\times 10^{-27} {\rm erg~cm^3\,s^{-1}}$. 
\end{tablenotes}

\end{threeparttable}
\end{table*}

\section{Clumpy Ultrafast Outflows} 
The high variability of UFOs offer a natural escape from these severe thermodynamic constraints that prevent the terminal velocity regions of outflows from entering a TI zone.  By introducing (month/year timescale) variability, it is no longer necessary to conclude that the only way to increase $\xi \propto F_{\rm X} v A$ without acceleration is for there to be a large divergence of the flow or an outward increase in the ionizing flux.  Rather, $\xi$ will increase according to $dF_{\rm X}/dt$. This appears to be a robust means for making supersonic outflows clumpy beyond their acceleration zones.  

The introduction of variability makes the problem as we pose it intrinsically time-dependent, yet as we find here, steady state \textit{unstable} background flow solutions can be obtained using a simple numerical setup.
Specifically, we perform 1D simulations using \athena \citep{Stone20} to solve, along the two different flowtubes shown in Fig.~\ref{fig:sketch}, the equations of non-adiabatic gas dynamics:
\beq
   \frac{D\rho}{Dt} = -\rho \nabla \cdot \gv{v}, \label{eq:cty}
\seq
\beq
   \rho \frac{D\gv{v}}{Dt} = - \nabla p - \rho \nabla \Phi_{\rm eff},
\seq
\beq
   \rho \frac{D\mathcal{E}}{Dt} = -p \nabla \cdot \gv{v} - \rho \mathcal{L}.
   \label{eq:energy}
\seq
Solutions for the RW and VW cases are found by running the code in spherical and cartesian coordinates, respectively.
The RW flowtube has $\gv{v} = v\,\hat{r}$, while the vertical wind (VW) flowtube has $\gv{v} = v\,\hat{z}$.
Besides this coordinate system change (which implicitly sets the flowtube area $A$), these flows are subject to different effective gravitational potentials;
for the RW case, $\Phi_{\rm eff} = -GM_{\rm bh}(1-\Gamma)/r$, and for the VW case, $\Phi_{\rm eff} = -GM_{\rm bh}(1-\Gamma)/\sqrt{x_0^2 + z^2}$.  From Fig.~\ref{fig:sketch}, both streamlines share a common foot point at $(x_0,z_0)$, as marked by a red dot; $(x_0,z_0)$ are set using $\delta = \tan^{-1}(z_0/x_0) = 25^\circ$.  

In \S{3.1}, we identify the free parameters of the problem to arrive at initial conditions appropriate for the environment of UFOs.
In \S{3.2}, we analyze the steady state background flow solutions obtained by raising the flux.  After discussing the numerical requirements to resolve the fastest growing TI modes in \S{3.3},
we present the results of introducing perturbations into these unstable solutions in \S{3.4} and \S{3.5}.  

Our treatment for perturbing the steady background flow solutions differs from \citetalias{Dannen20}: we apply a single \textit{spatial} perturbation episodically rather than a temporally periodic perturbation continuously (see \eqref{eq:delta_rho} below).  This permitted the comparison with linear theory given in \S{3.2} and \S{3.3}.
We found that \athena has difficulty evolving \eqref{eq:cty}-\eqref{eq:energy} using higher order methods once a perturbation is introduced.
Specifically, the code would crash when using the piecewise parabolic reconstruction method in combination with $\tt{integrator=rk3}$ (the strong stability preserving variant of the Runge-Kutta scheme) and the Harten-Lax-van Leer-Contact (HLLC) Riemann solver unless the perturbation amplitude is kept under about $0.2\%$. 
Such amplitudes result in clouds with only moderate density contrasts ($\delta\rho/\rho_0 \approx 1-2$).  
We therefore present results using the default numerical algorithms (van Leer integrator, piecewise linear reconstruction applied to primitive variables, and HLLC), as this permits cloud formation with $\delta\rho/\rho_0 \approx 4-5$.   

\begin{figure*}
  \centering
  \includegraphics[width=\textwidth]{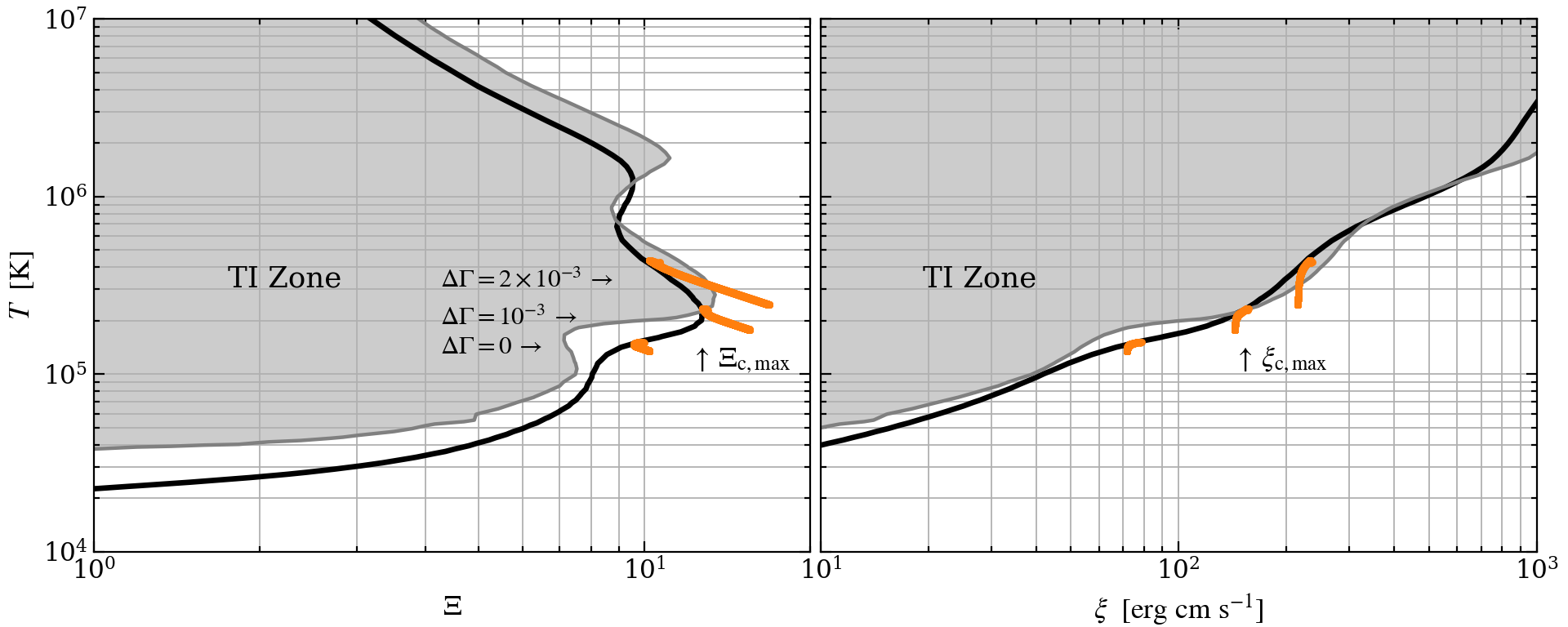}
  \caption{Phase diagrams illustrating how the background flow solutions of the RW flowtube can be made thermally unstable by increasing the radiation flux.  The bottom orange curve is the RW solution from Fig.~\ref{fig:A0}.
  The other two solutions start from the same initial conditions but evolve in response to an increasing luminosity given by \eqref{eq:ramp}. The $\Delta\Gamma = 10^{-3}$ ($\Delta\Gamma = 2\times 10^{-3}$) solution corresponds to a doubling (tripling) of the flux, which places the base (a large radial range) of the outflow into the TI zone.  
  The constant horizontal width of the `tracks' on the $(T,\xi)$ phase diagram (right panel) show that the amount of adiabatic cooling is unchanged (see the text and \eqref{eq:lambda_xi}), revealing that heat advection alone widens the tracks vertically.
  }
  \label{fig:DeltaGammas}
\end{figure*}

\subsection{Governing parameters}
Our supersonic outflow models are governed by four dimensionless parameters (to be defined presently): $\Gamma(t)$, $\Xi_0$, ${\rm HEP}_0$, and $\mathcal{M}_0$.\footnote{The subscript `0' denotes a quantity evaluated at $r_0$, except for $\Gamma_0$, where it denotes the initial Eddington fraction--- see \eqref{eq:ramp}.}
The thermally driven clumpy outflow solutions obtained by \citetalias{Dannen20} were self-consistently launched, meaning that the flow velocity $\gv{v}(\gv{r})$ is determined without specifying an `injection' velocity, and those solutions depended only on $\Gamma_0$, $\Xi_0$, and ${\rm HEP}_0$.  By not focusing in this paper on a particular outflow driving mechanism, a launching velocity must be assumed, i.e. the solutions depend on an additional free parameter, $\mathcal{M}_0$, the Mach number at $r_0$.  The initial Eddington fraction, $\Gamma_0 = \Gamma(t=0) = L(t=0)/L_{\rm Edd}$ (with $L_{\rm Edd} = 4\pi G M_{\rm bh} \bar{m}c/\sigma_{\rm es}$, where $\sigma_{\rm es}$ is the electron scattering cross section) couples to $\Xi_0$, the value of the pressure ionization parameter at the base of the flow,
\beq \Xi_0 = \f{p_{\rm rad}}{p_0} = \f{F_X/c}{n_0\, k\, T_0} = \f{f_X \Gamma_0}{n_0\, \sigma_{\rm es} r_0} \f{\bar{m} c^2}{k\,T_0} \f{r_g}{r_0}. \label{eq:Xi0} \seq
Here, the X-ray luminosity $L_X$ is specified from the total luminosity $L$ using $L_X = f_X \Gamma_0 L_{\rm Edd}$.
The location of the flow base, $r_0$, is set by the 
so-called hydrodynamic escape parameter (HEP), defined as 
$\rm HEP = |\Phi_{\rm eff}|/c_s^2$,
where $c_s = \sqrt{\gamma k\,T/\bar{m}}$ is the adiabatic sound speed (and we take $\gamma = 5/3$).  Evaluating the HEP at $r_0$ gives
\beq {\rm HEP}_0 =  3.92\times 10^7 \left(\f{r_0}{r_g}\right)^{-1}\left(\f{T_0}{10^5\, {\rm K}}\right)^{-1}(1 - \Gamma_0). \label{eq:HEP0}\seq

We briefly summarize how an outflow model can be constructed for a particular system; for a full description of the modeling framework, see \citetalias{Waters21}. The initial luminosity sets $\Gamma_0$, while the ionizing portion of the SED fixes $f_X$ (for our model SED, $f_X = 0.36$). The SED corresponds to a unique S-curve, which is used to set $T_0$ through $\Xi_0$, the location of the wind base on the S-curve.  
Given $\Gamma_0$, $\Xi_0$, and the distance $r_0$ as determined from a choice of ${\rm HEP}_0$, \eqref{eq:Xi0} is used to set the base number density, $n_0$.
Finally, the launching velocity $v_0$ is specified through a choice of $\mathcal{M}_0 = v_0/c_s(r_0)$.  While $\Gamma_0$, $\Xi_0$, ${\rm HEP}_0$, and $\mathcal{M}_0$ are free parameters, they obey basic constraints, as described in the Appendix.

\subsection{Unstable background flow solutions} 
\label{sec:background_flows}
In Table~1, we list values for the parameters of the simulation presented below, which were chosen to give an outflow with $\mathcal{M}_0 = 250$ at a distance $r_0 \approx 5\times 10^3\,r_g$.  The corresponding injection speed is $v_0 = 0.0495\,c$.  Given this choice, the initial conditions are set as $\rho(t=0) = \bar{m}n_0$ with $n_0$ found from \eqref{eq:Xi0}; $v(t=0) = v_0$; and $p(t=0) = n_0\,k\,T_0$.  

While the standard prescription for velocity injection is to set the desired value in the ghost zones, we found that this procedure leads to failed root solves within our semi-implicit heating and cooling routine.  Mass and velocity injection are instead applied to the first active zone.  In \athena notation, we fix $\tt{cons(IDN,k,j,is})$ to $\rho_0$, $\tt{cons(IM1,k,j,is})$ to $\rho_0\,v_0$, and $\tt{cons(IEN,k,j,is})$ to $\rho_0 \mathcal{E} + (1/2)\rho_0\,v_0^2$, where $\mathcal{E}$ is allowed to `float' (i.e. is determined by the semi-implicit scheme for net cooling, details of which can be found in the appendix of \citet{Dyda17}).  This procedure is then compatible with ghost zones being set by the default outflow boundary conditions, which are applied at both the inlet and outlet of the flowtube.  

Steady state solutions that enter the TI zone are obtained by first setting $n_0$ and $T_0$ on a cold branch of the S-curve, chosen here to be at $\xi_0 = 80$.  We then increase the radiation flux enough for part of the flow to enter the TI zone.
For simplicity, we increase the total luminosity rather than only the ionizing fraction $f_{\rm X}$, the ramp-up being done linearly in time using
\beq 
\Gamma(t) = \Gamma_0 + \Delta\Gamma\f{t}{t_{\rm rise}} ,
\label{eq:ramp}
\seq
with $\Gamma(t) = \Gamma_0 + \Delta\Gamma$ for $t > t_{\rm rise}$.
Since $\xi_{\rm c,max} \approx 141.3$, entering the TI zone requires $\Delta\Gamma/\Gamma_0 \gtrsim 1$.  A succession of quasi-steady solutions is expected when the ramp-up timescale satisfies $t_{\rm rise} > t_{\rm th}$, and this is what we observe.  We also find that $\xi$ can be increased when $t_{\rm rise} < t_{\rm th}$, just with a stronger transient passing through the profiles before a steady state can be reached. We set $t_{\rm rise} = 5\times10^7~{\rm s}$, corresponding to $t_{\rm rise}/t_{\rm dyn} \approx 1.9$ and $t_{\rm rise}/t_{\rm th} \approx 4.9$ (0.16) for the RW (VW) flowtube at $r_0$.  In Fig.~\ref{fig:DeltaGammas}, we plot the stable RW solution already analyzed in \S{2.1} along with the unstable steady state solutions obtained for $\Delta\Gamma/\Gamma_0 = 1$ and $\Delta\Gamma/\Gamma_0 = 2$.  

Notice the effect of a rise in $\Gamma(t)$: 
the vertical extent of the tracks on either phase diagram 
increases, so that while the flow at small radii has entered the TI zone, at some radius it also exits the TI zone.  
In light of our earlier discussion in \S{2.1}, it is interesting to assess whether this is due to
adiabatic cooling or heat advection.
The cause can readily be understood from \eqref{eq:heat3}.  The vertical height of the tracks in $\log T$-space is a direct measure of $\lambda_T$, whereas the horizontal width on the $(T,\xi)$-plane is only an indirect measure of $\lambda_\rho$ because, in the case of irradiation by a point source,
\beq 
\f{1}{\lambda_\xi} \equiv \nabla \ln \xi = -\f{2}{r} - \f{1}{\lambda_\rho}. \label{eq:lambda_xi}\seq
Nevertheless, the fact that the tracks only spread out vertically on the $(T,\xi)$-plane as $\Gamma(t)$ rises indicates that the rate of adiabatic cooling remains constant, while the rate of heat advection increases.  This behavior is expected given that $t_{\rm th} \propto T$, so $t_{\rm dyn}/t_{\rm th}$ decreases as the temperature rises upon climbing the S-curve.  Since initially $\lambda_\rho \approx -r/2$, i.e. is constant at any given location in the flow, we see from \eqref{eq:heat3} that a decrease in $t_{\rm dyn}/t_{\rm th}$ must be accompanied by an increase in $r_0/\lambda_T$.  That is, heat advection, not adiabatic cooling, is responsible for making the flow exit the TI zone.  

\begin{figure*}
  \centering
  \includegraphics[width=0.482\textwidth]{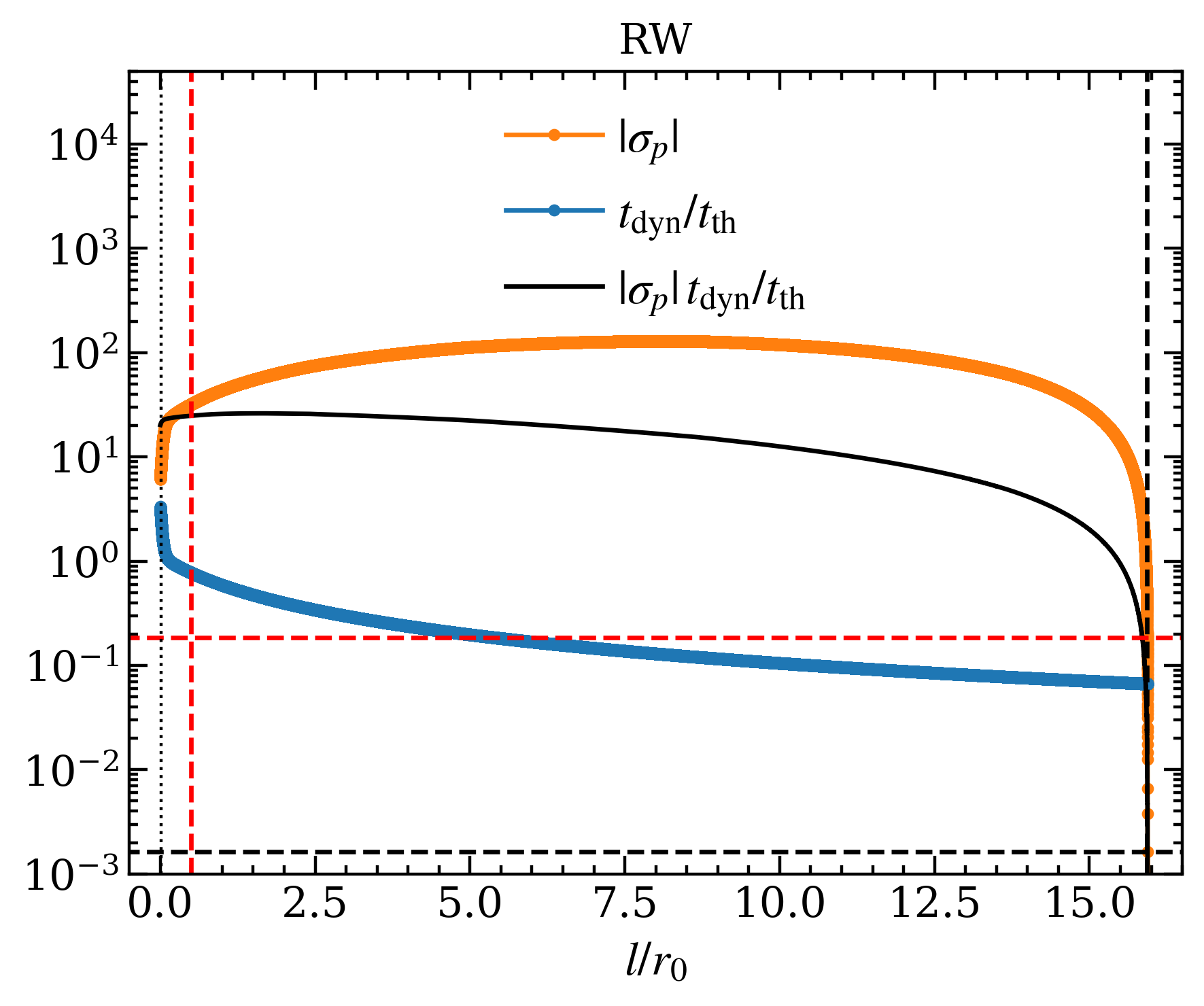}
  \includegraphics[width=0.448\textwidth]{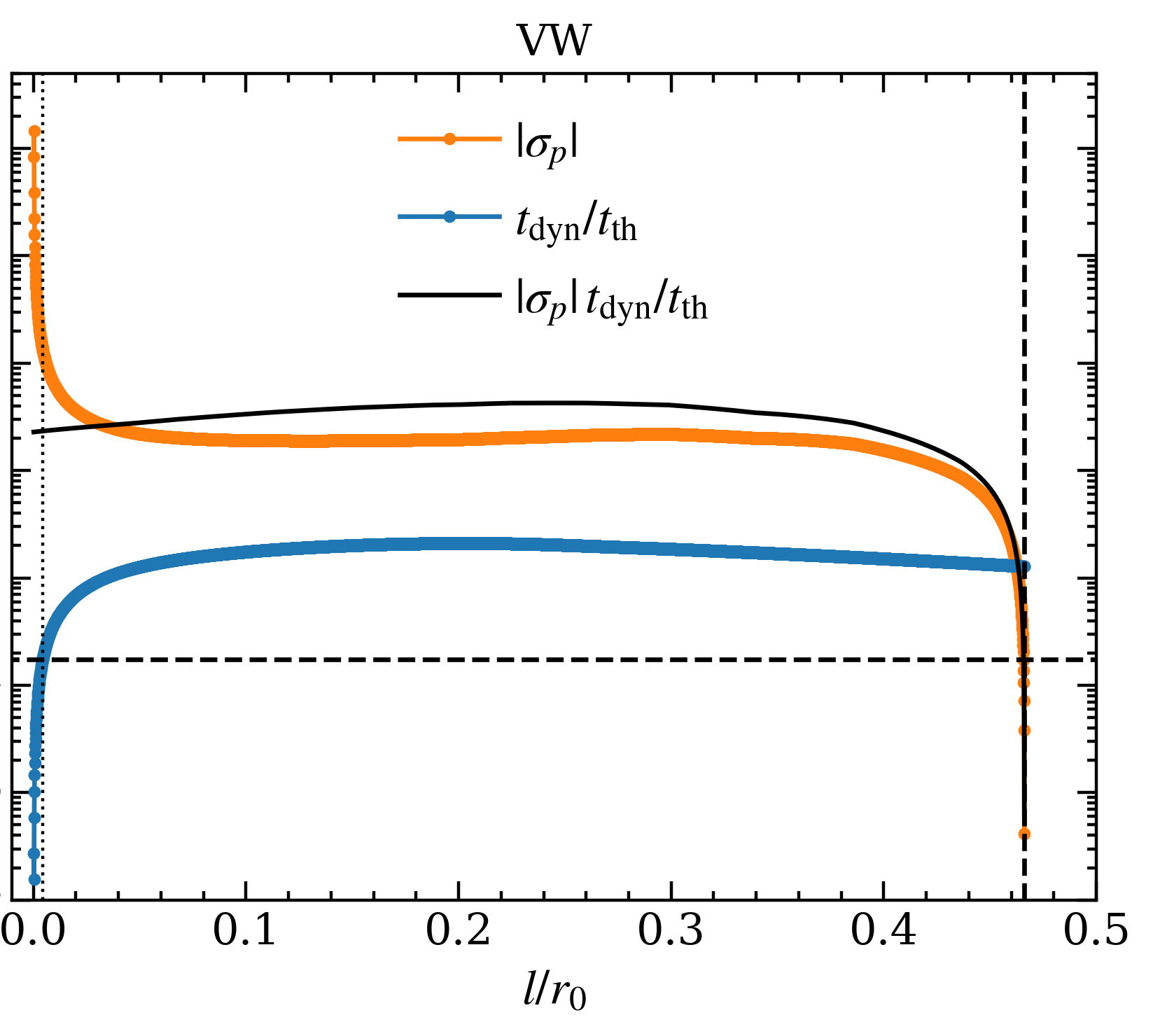}
  \caption{
  Stability analysis to assess whether or not (nearly) constant velocity flow along a streamline passing through a TI zone will actually permit cloud formation.  The vertical dashed lines mark the upper limit used in the integral of \eqref{eq:Delta_l}; black vertical lines are at the outer edges of the TI zones, while the red vertical line in the left panel shows a calculation where this upper limit is at $l=0.5\,r_0$ (to make a direct comparison with the VW case, for which the TI zone only extends to $l \approx 0.5\,r_0$).
  Horizontal lines mark the corresponding bounds found by evaluating \eqref{eq:tscale_bound}.  See the text for a description of why the intersection of the red lines constrains the `saturation' distance ${\Delta l}_{\rm sat}$, the location where a cloud will form once a perturbation is inserted at $l_p$ (marked here by the dotted black vertical lines near the flow base).
  }
  \label{fig:sigma_trat}
\end{figure*}

Having obtained steady state solutions partially occupying the TI zone, we can now address the problem of stability.  In Fig.~\ref{fig:sigma_trat}, we apply the theoretical analysis summarized by \eqref{eq:Delta_l} and \eqref{eq:tscale_bound}.  
In the right panel, the tendency for $\sigma_p$ and $t_{\rm th}$ to both become very large (making $t_{\rm dyn}/t_{\rm th}$ small) is due to their inverse dependence on $|L_{\rm net}|$ (see \eqref{eq:sigma_p} and \eqref{eq:t_th}), which rapidly approaches 0 for near-equilibrium positions on the S-curve.  However, this is not truly diverging behavior because a non-adiabatic flow cannot exactly occupy the S-curve (i.e. have $|L_{\rm net}| = 0$), for this would violate the heat equation (see \eqref{eq:heat}).  The growth rate of entropy modes depends only on the product $\sigma_p\,t_{\rm dyn}/t_{\rm th}$, which is shown by the well-behaved black lines.

Recall that ${\Delta l}_{\rm TI, max}$ in \eqref{eq:Delta_l} measures how far into the TI zone an entropy mode would have to travel before encountering the maximum growth rate.  It is found by integrating over the orange curves in Fig.~\ref{fig:sigma_trat} and dividing by $|\sigma_p|_{\rm max}$, the peak value along these curves.  These values are ${\Delta l}_{\rm TI} = 10.9\,r_0$ for the RW solution and ${\Delta l}_{\rm TI} = 9.0\times 10^{-4}\,r_0$ for the VW solution.  Evaluating the right hand side of \eqref{eq:tscale_bound} over the entire TI zone then provides a lower limit on $t_{\rm dyn}/t_{\rm th}$, which is marked by the dashed horizontal black lines.  Since this bound is well satisfied, we can conclude that these solutions will allow dynamical TI to saturate, thereby providing clumpy outflow models of UFOs.

It is also possible to graphically estimate the `saturation distance' of dynamical TI, ${\Delta l}_{\rm sat}$, i.e. the distance at which a clump will first appear.  
In the VW case, we see that $t_{\rm dyn}/t_{\rm th}$ rises above the black horizontal line very close to the base of the flow.  Due to the growth rate being very large and peaking at the base, we have ${\Delta l}_{\rm TI} \approx {\Delta l}_{\rm sat}$.  To utilize this bound in the RW case, we evaluate ${\Delta l}_{\rm TI}$ over just a portion of the TI zone to assess when it is first satisfied. 
Integrating only up to $l=0.5\,r_0$, as marked by the red vertical line, puts ${\Delta l}_{\rm sat}$ at the intersection with the red horizontal line, where \eqref{eq:tscale_bound} is still well satisfied. Thus, we see that ${\Delta l}_{\rm sat}$ in the RW case is also very close to $r_0$.

The foregoing analysis is of general applicability since even time-dependent flows typically have well defined streamlines.  Whenever a streamline occupies a TI zone, the flow is formally thermally unstable, but the exponential growth of isobaric entropy modes can only lead to the saturation of dynamical TI if $t_{\rm dyn}/t_{\rm th}$ satisfies the bound given in \eqref{eq:tscale_bound}. 

\subsection{Resolution requirements}
As we now derive, the number of grid zones needed to resolve the fastest growing modes of TI scales as $\mathcal{M}_0^2$, so we state the lesson learned from this analysis up front: 
a typical outcome of high resolution UFO simulations may end up being that the outflow cannot become clumpy, even when the bound in \eqref{eq:tscale_bound} is well satisfied, because this requires \textit{extremely high} rather than just high resolution.
Stated differently, while it is computationally inexpensive to obtain steady state solutions with speeds $v/c \gtrsim 0.01$, it can become numerically prohibitive to arrive at time-dependent, clumpy UFO solutions --- even in 2D.
By comparison, our 1D clumpy outflow solutions in the warm absorber parameter regime were found using about $8,000$ grid zones \citepalias[see][]{Dannen20}, 50 times fewer than needed here.  The gas density was orders of magnitude smaller in those solutions, so the clouds were much larger.  Isobaric cloud sizes are on the order of the thermal length scale, 
\beq \lambda_{\rm th} \equiv c_s t_{\rm th}, \seq
termed the coherence length by \citet{PD85},
and this is much easier to resolve in warm absorber outflows because for a fixed temperature, $\lambda_{\rm th}\propto \rho^{-1}$.

\begin{figure}
  \centering
  \includegraphics[width=0.49\textwidth]{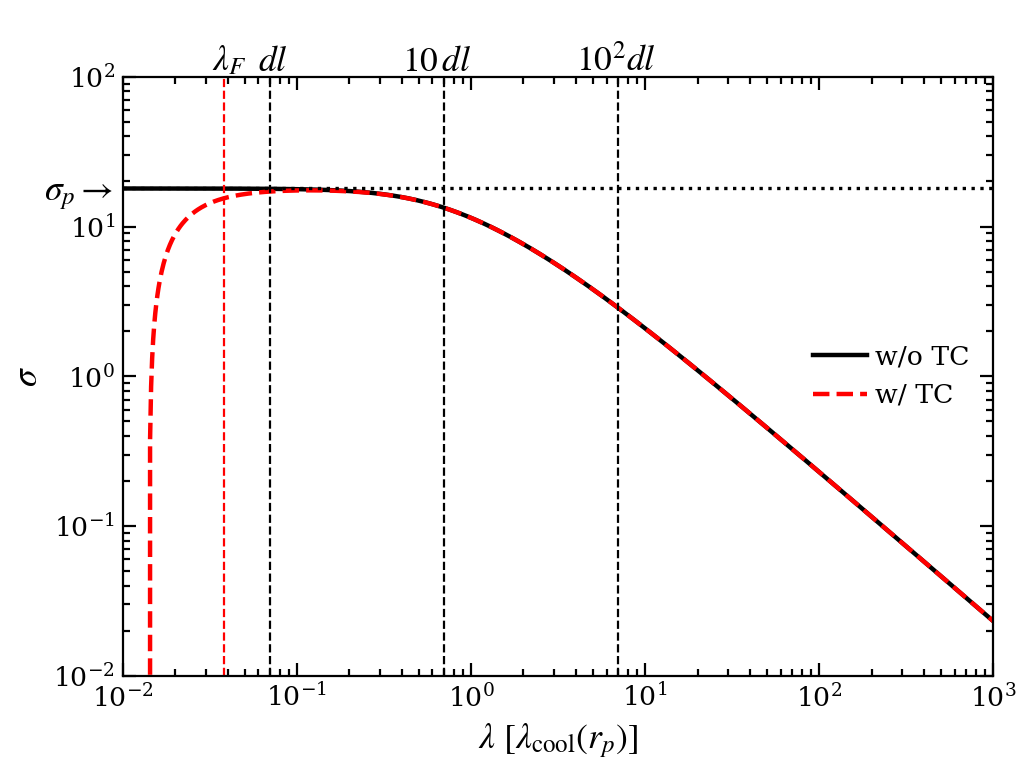}
  \caption{Solution to Field's cubic dispersion relation for local TI, with (red dashed line) and without (black solid line) including thermal conduction (TC), evaluated at $r_p=1.01\,r_0$ for the RW flowtube using the parameters in Table~1.  The horizontal dotted line marks $\sigma_p$, the isobaric growth rate appearing in the integrand of the \citet{Mathews78} amplification solution given in \eqref{eq:growth2}.  The only growth rates that can be picked up in grid-based numerical calculations are ones with wavelengths $\lambda > 4\,dl$; the grid spacing $dl$ is marked by the leftmost vertical black line (labels at the top are also in units of $\lambda_{\rm cool}(r_p)$).  All resolvable growth rates therefore exceed the Field length, $\lambda_F$, showing that including TC will not alter the growth rates.  The spatial perturbations introduced by hand in our calculations have a wavelength $\lambda = 10\, dl$, so the fastest growth rate sampled is $\sigma \approx 0.8\,\sigma_p$ (see middle vertical black line).  The rightmost vertical line marks the fastest growth rate that can be resolved in runs with $N_l \approx 41,000$ zones instead of $N_l \approx410,000$ zones.  
  }
  \label{fig:DR}
\end{figure}

The \citet{Mathews78} solution in \eqref{eq:growth} applies to locally isobaric entropy modes.  These modes will have a wavelength 
$\lambda \sim \lambda_{\rm cool} = (|\mathcal{L}|/\Lambda_0) \lambda_{\rm th}$ (see \eqref{eq:t_th})
evaluated in the background flow \citep{Waters19a}.
To estimate the resolution required to resolve these modes, notice that because $\lambda_{\rm th} \propto n^{-1}T^{3/2}$, it is sufficient 
to know $\lambda_{\rm th}$ at $r_0$ and the behavior of the profile of $n^{-1}T^{3/2}$ along a streamline.  For the RW flowtube, $n^{-1}T^{3/2}$ is a strongly increasing function of radius, so the shortest wavelength modes will be those at $r_0$.  Using \eqref{eq:Xi0} and \eqref{eq:HEP0} to express the density dependence of $t_{\rm th}(r_0)$ in terms of the governing dimensionless parameters gives
\beq 
\begin{split}
\f{\lambda_{\rm th,0}}{r_0} =  \f{3.96 \times 10^{-4}}{f_X}
\left|\f{L_{\rm net}}{10^{-24}}\right|^{-1}
&\left(\f{\Gamma}{10^{-3}}\right)^{-1}
\left(\f{{\rm HEP}_0}{10^3}\right)^{-1} \\
& \left(\f{\Xi_0}{10}\right)
\left(\f{T_0}{10^5\, {\rm K}}\right)^{\f{3}{2}}.
\end{split}
\label{eq:lambda_th}
\seq
Here we have taken $1-\Gamma \approx 1$ (see \eqref{eq:Gamma_bound}). 
Since the resulting clumps can be as small as a single entropy mode, the resolution per thermal length scale will be comparable to the initial resolution within clumps.  Requiring that $\lambda_{\rm th,0}$ be resolved by at least $10$ zones (i.e. for a grid spacing $dl < \lambda_{\rm th,0}/10$), and defining the number of grid zones as $N_l = l_{\rm out}/dl = D\, r_0/dl$ (for dynamic range $D =l_{\rm out}/r_0$), 
we find $N_l > 10\,D\, r_0/\lambda_{\rm th,0}$, or
\beq 
\begin{split}
N_l \gtrsim  10^{6} f_X\,
\left|\f{L_{\rm net}}{10^{-24}}\right|
\left(\f{D}{40}\right) 
&\left(\f{\Gamma}{10^{-3}}\right) 
\left(\f{{\rm HEP}_0}{10^3}\right) \\
&\left(\f{\Xi_0}{10}\right)^{-1}
\left(\f{T_0}{10^5\, {\rm K}}\right)^{-\f{3}{2}}.
\end{split}
\label{eq:Nlbound}
\seq
By \eqref{eq:M0bound}, we have arrived at the above noted scaling $N_l \propto \mathcal{M}_0^2$. This result follows from having $n_0 \propto r_0^{-2}$ for a constant $\Xi_0$ by \eqref{eq:Xi0} and is the statement that if the flow velocity is comparable to the local escape speed, then cooling times become shorter and thus isobaric wavelengths become harder to resolve as the escape speed increases at smaller $r_0$.  

Plugging in the parameters from the RW column in Table~1 gives $N_l \approx 32,000$ for $D=40$ (recall that $f_{\rm X} = 0.36$).
However, a precise calculation of the maximum growth rate reveals that $\sigma_p$ is reached at wavelengths about five times smaller than $\lambda_{\rm cool}(r_0)$.  Hence, resolving the fastest isobaric modes actually requires $N_l \gtrsim 1.6\times 10^5$ zones, which would make even 2D simulations at this resolution a very resource-intensive problem on current supercomputers.  This calculation is shown in Fig.~\ref{fig:DR}, where we plot solutions to the cubic dispersion relation of \citet{Field65} with and without thermal conduction.  We mark our grid spacing with a vertical line labeled `$dl$'.  Well resolved perturbations will have growth rates corresponding to $\lambda \gtrsim 10\, dl$, so $N_l$ was chosen to place $10\,dl$, the wavelength of the spatial modes we introduce into the background flow solutions, at the location where $\sigma \approx \sigma_p$.  To simultaneously show that the clouds which can form are not subject to classical evaporation and that convergence can in principle be reached using our numerical setup, we also calculated the Field length $\lambda_F$ \citep[][see the red vertical line]{Begelman90}. 
The red and black curves overlap for $\lambda > 0.1\, \lambda_{\rm cool}$ indicating that including thermal conduction would not alter the maximum growth rate and would have no effect on the perturbations we introduce by hand.  

\begin{figure*}
  \centering
  \includegraphics[width=0.445\textwidth]{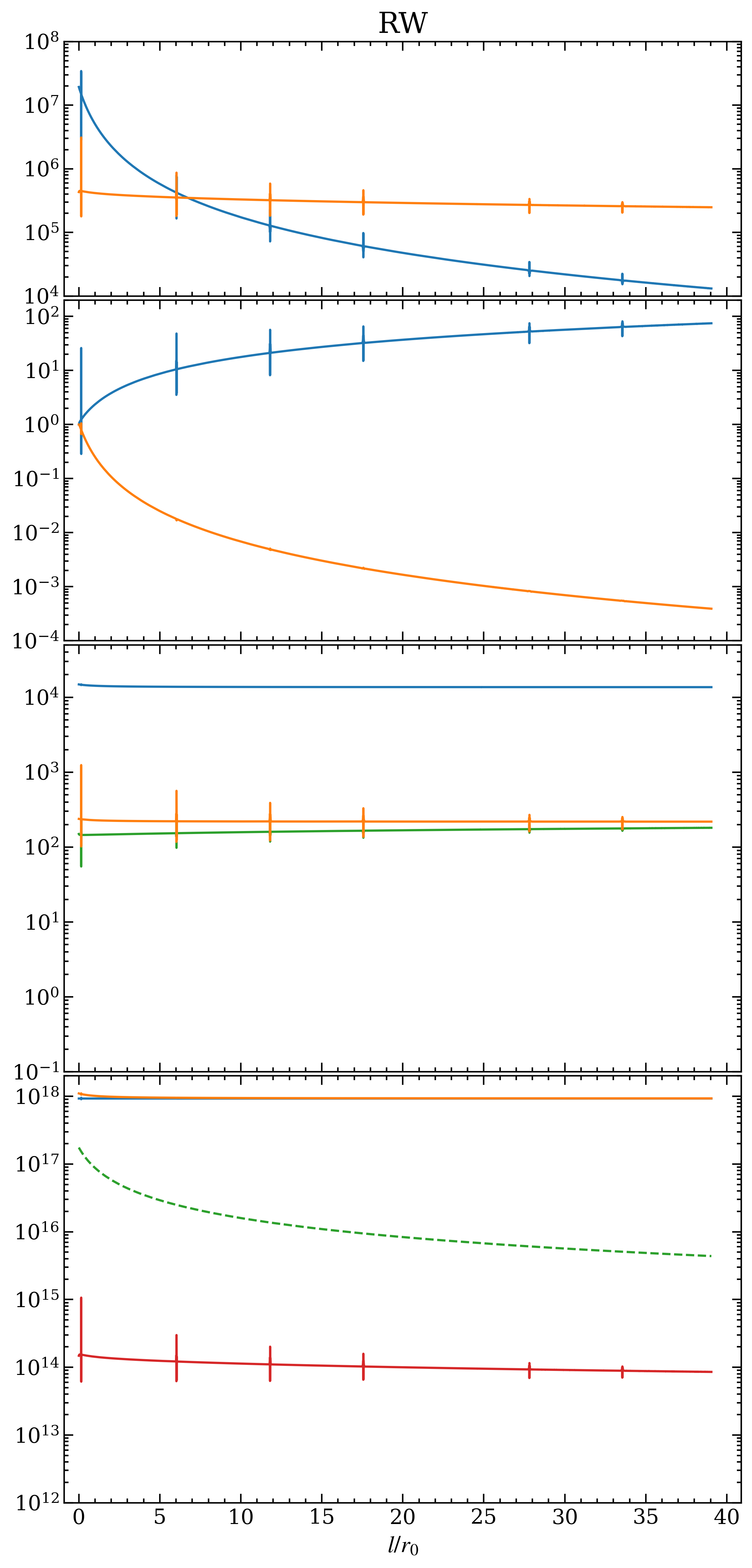}
  \includegraphics[width=0.41\textwidth]{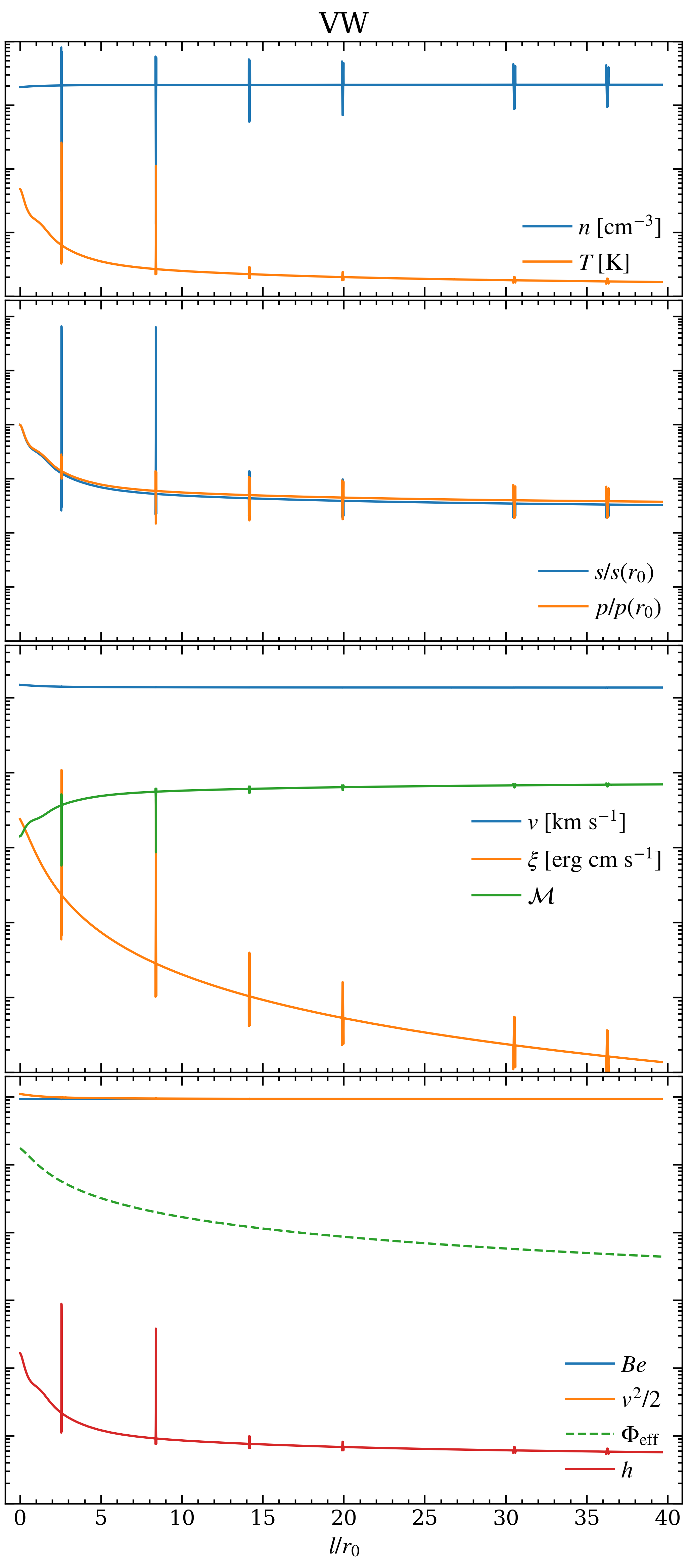}
  \caption{Profiles of various quantities along a streamline for the RW flowtube (left panels) and the VW flowtube (right panels) after introducing perturbations into the background flow solutions.  
  The legends apply to both columns.  In the bottom panels, $h = c_s^2/(\gamma-1)$ is the enthalpy and the dashed line indicates that $\Phi_{\rm eff} < 0$.  Notice the clouds are only visible as `spikes' on top of the background flow solutions because their characteristic size is $\lambda_{\rm th} \sim 10^{14}\,{\rm cm} \approx 10^{-2}r_0$ --- see \eqref{eq:lambda_th}.  The clouds are nevertheless fully resolved (see Fig.~\ref{fig:clumps}).  Solutions shown here are for a representative snapshot; an animated version of this plot can be viewed at \url{https://trwaters.github.io/dynamicalTI/}.
  }
  \label{fig:profiles}
\end{figure*}

\begin{figure*}[ht]
  \centering
  \includegraphics[width=\textwidth]{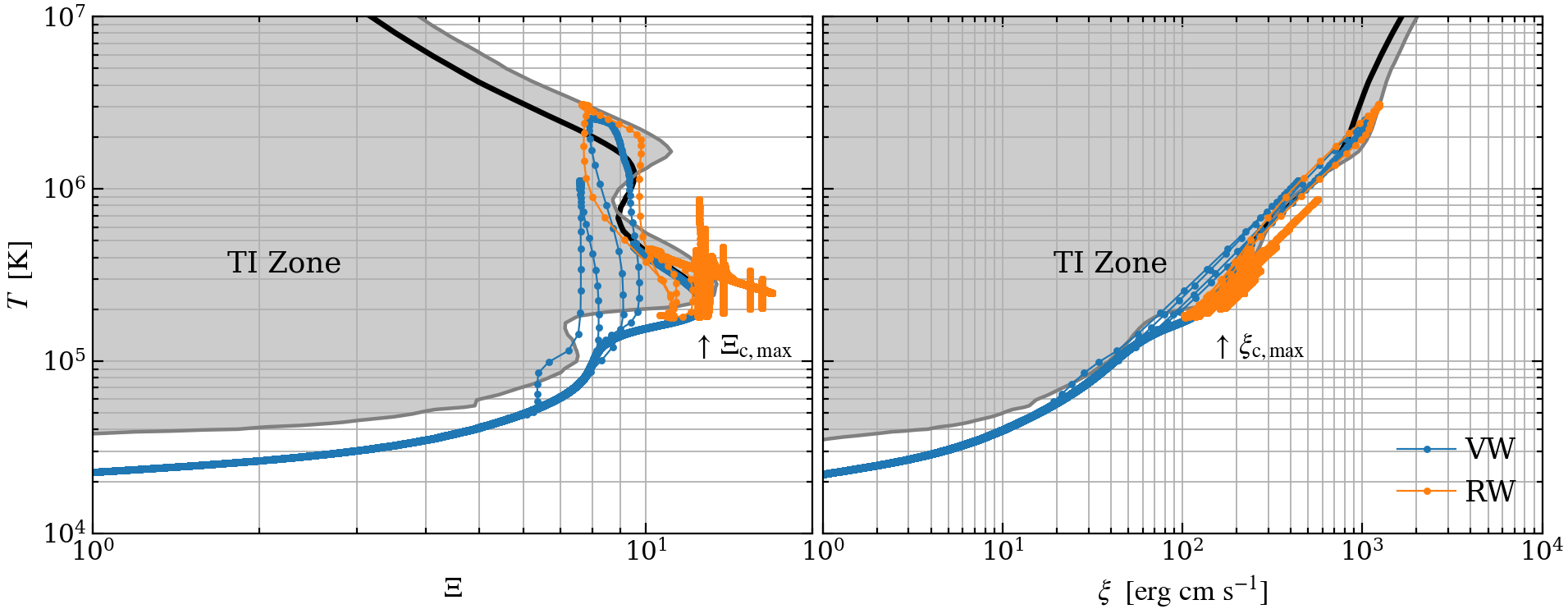}
  \caption{`Tracks' of the clumpy UFO solutions plotted in Fig.~\ref{fig:profiles} shown on either phase diagram, as in Fig.~\ref{fig:A0}.  The RW tracks (orange) correspond to the $\Delta \Gamma = 2\times 10^{-3}$ solution shown in Fig.~\ref{fig:DeltaGammas} upon applying perturbations.  Symbols mark grid zone locations to give a sense of the numerical resolution within a cloud. An animated version is this plot can be viewed at \url{https://trwaters.github.io/dynamicalTI/}. 
  }
  \label{fig:Ak1e-2}
\end{figure*}

\subsection{Clumpy UFO solutions}
We now explore the nonlinear regime of dynamical TI by perturbing the background flow solutions from \S{\ref{sec:background_flows}}. 
By design, our procedure to generate entropy modes is resolution-dependent.  We introduce a spatial perturbation of wavelength $\lambda = 10\, dl$ and amplitude $a_0 = 10^{-2}$ just above the base of the flow at $r_p = 1.01\,r_0$. Specifically, we apply the following perturbation to the density field every $6.2\,t_{\rm dyn}$:
\beq 
\delta \rho = 
\begin{cases}
    a_0\, \rho_0 \sin\left(\f{2\pi r}{\lambda}\right) ,& \text{if } r_p - \f{\lambda}{2} < r < r_p + \f{\lambda}{2}\\
    0,              & \text{otherwise.}
\end{cases}
\label{eq:delta_rho}
\seq 
For already high resolution runs with $N_l = 40,992$, these perturbations have almost no noticeable effect on the solutions. The growth rate for this size wavelength is marked by the rightmost vertical line in Fig.~\ref{fig:DR} and is $\sigma \approx \sigma_p/35$; the level of amplification at this rate is observed to be about $20\%$, corresponding to a tiny `density bump' advecting downstream.  We therefore present results for $dl$ set using $N_l = 409,920$, corresponding to $dl = 3.72\times 10^{12}\,{\rm cm} \approx (0.48\,r_g)$, as only these near-maximum growth rates (see the vertical line at $10\, dl$ in Fig.~\ref{fig:DR}) are sufficient to cause clump formation.  

The resulting solution profiles are shown in Fig.~\ref{fig:profiles}.
In the bottom panels, we plot a breakdown of the Bernoulli function, $Be = v^2/2 + h + \Phi_{\rm eff}$ (with $h = c_s^2/(\gamma-1)$ the enthalpy), to give a sense of the overall energetics. The enthalpy profiles reveal that pressure effects still regulate the gas density, even for a flow that is kinetic energy dominated.

For the RW flowtube (left panels), the largest density in the domain occurs in the innermost clump (here visible as just a `spike' since $\lambda \ll r_0$).  Consistent with our analysis in \S{3.2}, this indicates that the saturation of TI is almost immediate, occurring over a distance $\Delta l \ll r_0$.  Spikes are not visible in the pressure profile of the RW solution, and this is the expected property of cloud formation in the isobaric regime \citep[see][]{Waters19a}.   
The constancy of the bulk velocity profile (and hence of the kinetic energy profile; see the lower two panels of Fig.~\ref{fig:profiles}) 
can also be understood from linear theory, as condensation dynamics in the comoving frame of the cloud is subsonic.
Because neither velocity profile features spikes, the kinetic power of the outflow, $P_k = (1/2)\dot{M} v^2$, can only be enhanced by the density contrast with respect to the background flow.
However, the pressure profile of the VW solution does show spikes, which is important to understand since this can alter the shock dynamics once this wind plows into an ambient medium.
We return to analyzing this non-isobaric behavior in \S{3.5}.

\begin{figure*}[ht]
  \centering
  \includegraphics[width=0.95\textwidth]{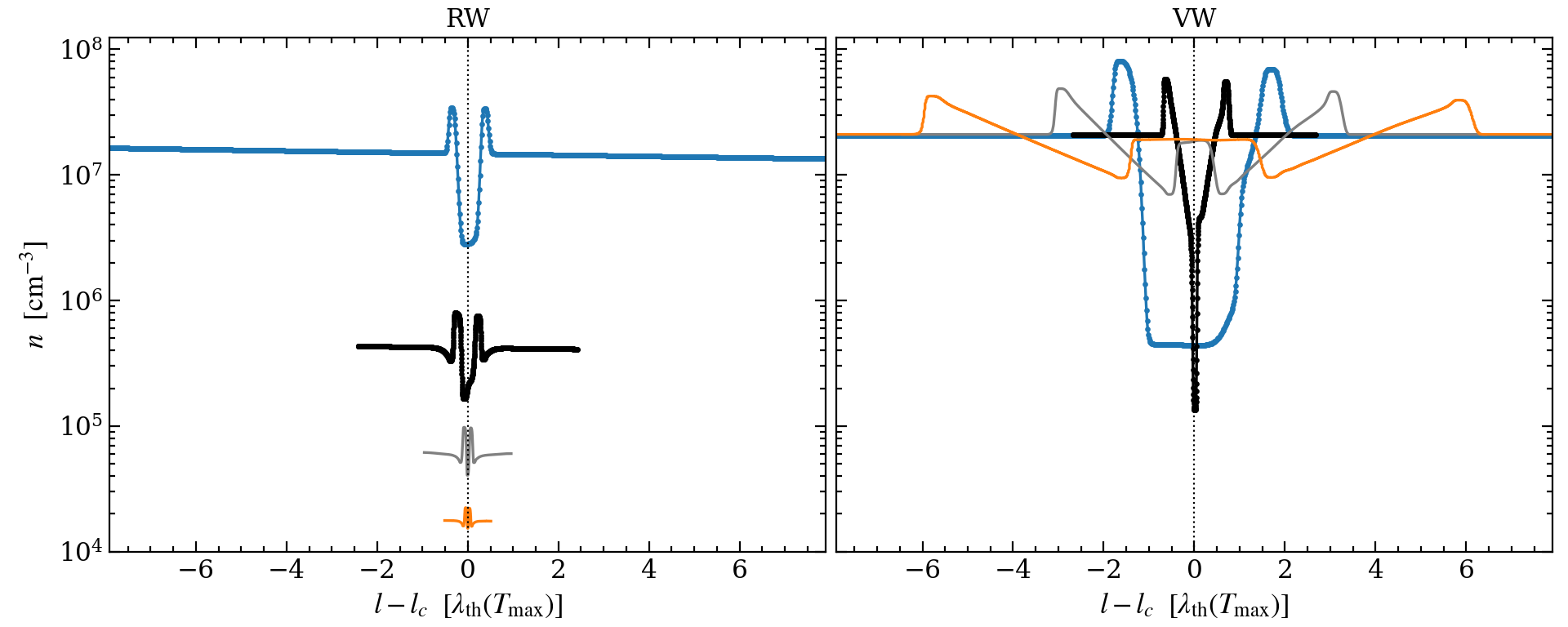}
  \includegraphics[width=0.95\textwidth]{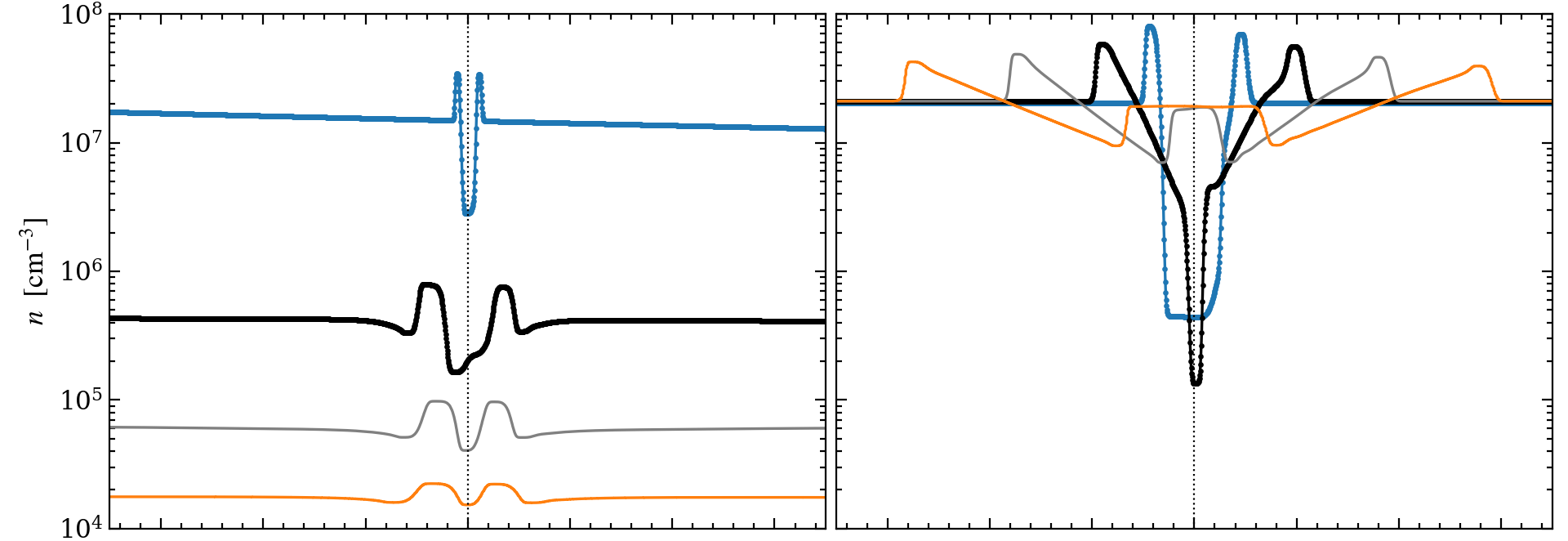}
  \includegraphics[width=0.95\textwidth]{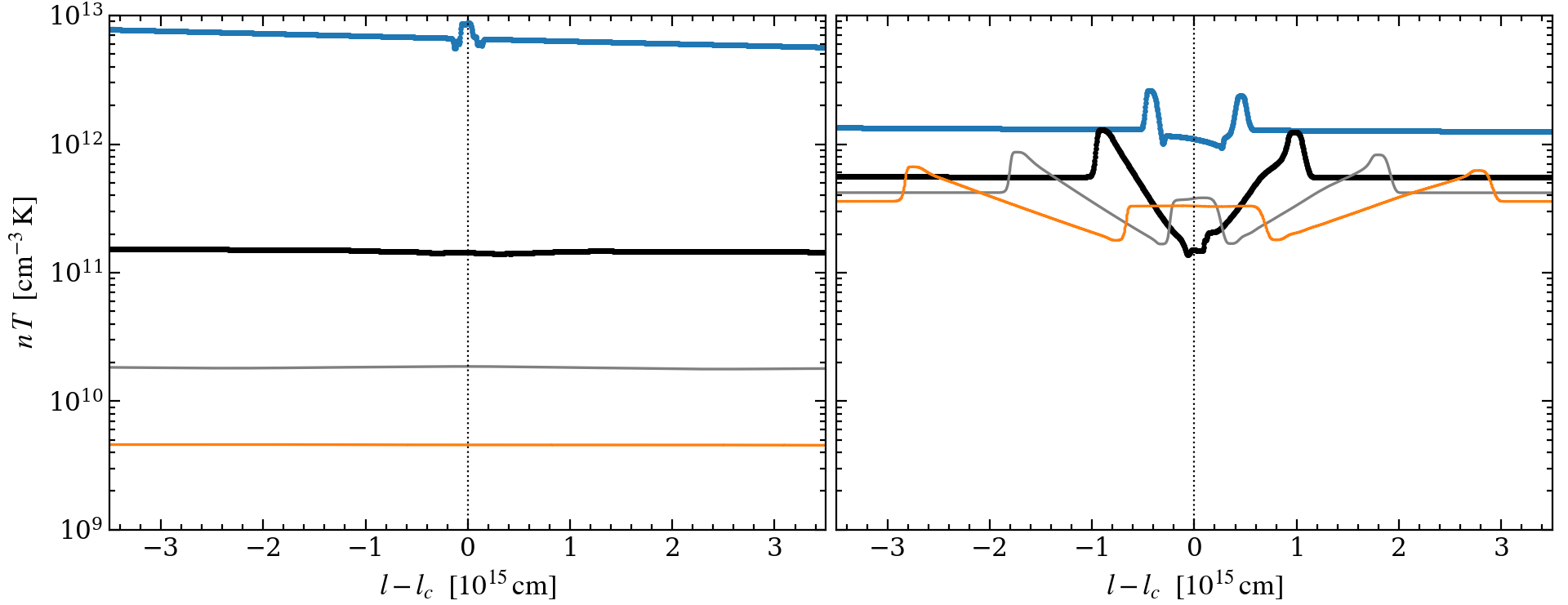}
  \caption{Close-up analysis of the clouds, showing that RW clumps are isobaric, while VW clumps are non-isobaric.  
  Blue, black, gray, and orange curves are zoom-ins on the 1st, 2nd, 4th, and 6th `spikes' from Fig.~\ref{fig:profiles}; following a single cloud with time would show the same behavior.  Symbols on the blue and black curves denote the locations of grid zones; the clouds are thus well-resolved.
  On the $x$-axis label, $l_c$ denotes the distance to the center of each cloud. \textit{Top panel}: Number density profiles versus the local coordinate $l-l_c$ in units of $\lambda_{\rm th}(T_{\rm max})$, the thermal length evaluated at the maximum temperature over the plotted range (2,000 zones centered on the zone at $l_c$).  This length measures how far sound waves travel in a thermal time, so $\lambda_{\rm th}(T_{\rm max})$ corresponds to $\lambda_{\rm th}$ of the fastest sound waves.  
  \textit{Middle and bottom panels}: Density and pressure profiles but with $l-l_c$ in units of $10^{15}\rm{cm}$ to depict the evolution of clump sizes and pressure differences in physical units.}  
  \label{fig:clumps}
\end{figure*}

In Fig.~\ref{fig:Ak1e-2}, we compare these same RW and VW snapshots on either phase diagram.  This plot shows that there are major differences in the evolution of the clumps.  In the RW solution (orange), each clump is visible on the $(T,\Xi)$-plane, whereas only two clumps can be seen in the VW solution.  The reason is simply that the evolution of the clumps is dictated by the dynamics of the background flow solution.   
For the VW clumps to span a range of temperatures above the S-curve, the warm phase of the clump would have to continue to heat up until it reaches a stable temperature above the TI zone.  But this warm phase gas comoves with the cold phase gas that is being advected into a lower temperature region at low $\Xi$.  Because $t_{\rm dyn}/t_{\rm th} \lesssim 2$ by Fig.~\ref{fig:sigma_trat}, the rate of radiative heating cannot compete with that of advective cooling, and so the clump survives only as a condensation occupying the cold branch of the S-curve.  The temperature profile in Fig.~\ref{fig:profiles} shows this also; the range of temperatures of all but the first two clumps are very small.  The RW clumps also exit the TI zone due to the advective cooling taking place in the background flow, but because the reduction in temperature is much less, the clumps take significantly longer to `rejoin' the background flow.  

\subsection{Isobaric vs. non-isobaric evolution}
In the local TI approximation, the velocity and pressure of the condensing gas in the rest frame of the cloud at the time of saturation is $\sigma_p (k\, t_{\rm th})^{-1}$ and $\rho\, \sigma_p^2 (k\, t_{\rm th})^{-2}$, respectively, where $k=2\pi/\lambda$ is the wavenumber \citep{Waters19a}. Hence, both the kinetic and thermal energy of the saturation process are of the same order, with the latter equal to (after re-introducing $t_{\rm dyn}$),
\beq 
e_{\rm sat} = \left(\f{\lambda}{r_0} \f{\sigma_p}{2\pi} \f{t_{\rm dyn}}{t_{\rm th}} \right)^2 v^2
= \left(\mathcal{M} \f{\lambda}{r_0} \f{\sigma_p}{2\pi} \f{t_{\rm dyn}}{t_{\rm th}} \right)^2 c_s^2.
\label{eq:esat}
\seq
The second equality in terms of Mach number lets us diagnose when the pressure profile should reveal the presence of the clumps.  The saturation energy of the entropy mode is comparable to the thermal energy of the background flow when $e_{\rm sat}/c_s^2 \sim 1$.  Since $\lambda/r_0 \sim 10^{-3}$, this occurs when $\mathcal{M}\, \sigma_p\, t_{\rm dyn}/t_{\rm th} > 2\pi \times 10^3$.  For the RW flowtube, $\mathcal{M} \approx 150$ and, as shown by the black line in the left panel of Fig.~\ref{fig:sigma_trat}, $\sigma_p\,t_{\rm dyn}/t_{\rm th} < 30$.  For the VW flowtube, however, both quantities are larger (see the right panel of Fig.~\ref{fig:sigma_trat}) and the condition $e_{\rm sat}/c_s^2 > 1$ is satisfied.  This explains why the pressure profile shows spikes in this case; physically, the pressure cannot respond to the condensation dynamics taking place because the thermal time is not short enough.  

A closer examination of terms in the Bernoulli function further adds to this analysis (see the bottom panels in Fig.~\ref{fig:profiles}). In both flowtubes, $v^2/2 \gg |\Phi_{\rm eff}| \gg h$ and therefore $Be \approx constant$.  Consistent with the dominance of $v^2/2$ compared with $h$, the first equality in \eqref{eq:esat} shows that $e_{\rm sat}/v^2 \ll 1$, hence there are no spikes visible in either velocity profile.  Despite $h$ being dynamically unimportant globally, the absence of spikes in the RW pressure profiles indicates that the enthalpy is significantly larger for this flowtube, which is seen to be the case. 

We emphasize that we are able to apply linear theory results from the local TI approximation to global simulations only because the UFO parameter regime is just one step removed from local TI simulations.  By this we mean that the growth rates are determined from the solution to Field's cubic dispersion relation, as plotted in Fig.~\ref{fig:DR}.  
By contrast, in the parameter regime of warm absorbers that we explored in \citetalias{Dannen20}, the wavelength of isobaric perturbations are comparable to the density and temperature gradient scale lengths.  Entropy modes introduced into those solutions, in addition to sampling changing background flow conditions, have growth rates that are a priori unknown, being governed by a more complicated dispersion relation.  The requirement for the local TI dispersion relation to apply is for the wavelength of perturbations to be small compared to all background flow gradient scale lengths, and Fig.~\ref{fig:profiles} shows this is the case for isobaric perturbations.  

The degree of non-locality in these simulations is therefore determined by how fast the local isobaric growth rate changes as the perturbation is advected a distance $\min(\lambda_\rho,\lambda_\xi,\lambda_T)$, i.e. into a region with significantly different background parameters.  This, in turn, is governed by the value of $t_{\rm dyn}/t_{\rm th}$: if this ratio is large, the perturbation can be exponentially amplified to the point that TI saturates before ever sampling different background flow conditions. Fig.~\ref{fig:sigma_trat} shows that $t_{\rm dyn}/t_{\rm th}$ is instead \textit{small} at the time of saturation for VW clumps, indicating that a perturbation samples differing flow conditions before becoming nonlinear.  

An intuitive understanding of why RW and VW clumps must evolve differently
follows by first recalling that $t_{\rm th}$ measures how fast the gas temperature can respond to changes in the heating and cooling rates.  Physically, for a newly formed cloud to reach pressure equilibrium while continually being subjected to different rates as it flows outward into smaller temperature regions, the sound crossing time, $t_{\rm cross}$, must be less than $t_{\rm th}$.
It then becomes immediately clear why VW clumps tend toward non-isobaric evolution: $t_{\rm th}$ cannot decrease by much because the clump density changes very little, while $t_{\rm cross}$ will increase significantly due to the sharp decrease in temperature.    

To illustrate non-isobaric evolution and to ascertain the behavior of $t_{\rm cross}/t_{\rm th}$ for RW clumps, in Fig.~\ref{fig:clumps} we overplot and zoom-in on four of the clumps from Fig.~\ref{fig:profiles}.  Denoting the size of the cloud by $\lambda_c = c_s\,t_{\rm cross}$, the condition $t_{\rm cross}/t_{\rm th} < 1$ is equivalent to $\lambda_c/\lambda_{\rm th} < 1$.  In the top panel, we therefore plot the profiles of the clumps in units of $\lambda_{\rm th}$. The RW clumps, which become less dense with time because of the $1/r^2$ falloff in the background density, also become more isobaric with time, with $\lambda_c/\lambda_{\rm th}$ decreasing. Their physical size, meanwhile, increases due to the expansion of the flow, as shown in the middle panel.  
The VW clumps also increase in size with time, but the reason cannot be attributed to the background flow dynamics; it must instead be due to the internal dynamics driven by the non-isobaric pressure field.  Consistent with the above expectations, the top panels reveal that $\lambda_c/\lambda_{\rm th}$ increases with time for these clumps. 
Comparing the pressure profiles in the bottom panels of Fig.~\ref{fig:clumps}, we sum up this final result:
for clouds formed from initially isobaric perturbations, it is possible for the evolution to become increasingly non-isobaric with time.

\section{Discussion}
Non-adiabatic thermal winds, the class of ionized AGN outflow models 
that naturally include the physics of TI, were only recently demonstrated 
to be susceptible to becoming multiphase \citepalias{Dannen20}.  
The parameter space for dynamical TI to operate was shown to be quite narrow, 
corresponding to distances far beyond the Compton radius (parsec scales or greater).  
In this work, we investigated if, on much smaller scales, there is additional 
parameter space that simply had not yet been explored.
Specifically, here we considered the kinetic energy dominated regime, 
the energetics hierarchy being
$v^2/2 \gg |\Phi_{\rm eff}| \gg \mathcal{E}$.
While this solution regime is qualitatively different from that
explored by \citetalias{Dannen20}, a unifying principle becomes clear
upon noting the finding by \citetalias{Waters21} that a special case of Crocco's theorem 
is the basis for understanding when dynamical effects will stabilize the 
flow to TI.  This theorem applies to our steady state background flow solutions; along streamlines, it reduces to simply
$\nabla Be = T\nabla s$,
where $s$ is the entropy per unit mass.  Entropy modes are subject
to the disruptive effects of stretching terms (discussed in \S{2})
when the entropy profile features a large gradient in regions 
that satisfy Balbus' instability criterion
(i.e. when there is a kink in $s(l)$ within the TI zone).  

By Crocco's theorem then, dynamical TI will be suppressed 
whenever there is a large gradient in the profile of $Be$ within the TI zone.  
In the present solutions, $Be$ is nearly constant
due to the flow being kinetic energy dominated, hence entropy modes are 
permitted to grow uninterrupted.  
The connection with thermal wind theory can now be stated succinctly: 
in both the present solutions and the thermal wind solutions of \citetalias{Dannen20},
the most favorable conditions for dynamical TI are found in unbound regions 
of the flow; here, the wind is unbound by construction with $Be \approx constant > 0$,
whereas in \citetalias{Dannen20} it is unbound only 
beyond a critical distance corresponding to $Be=0$, as shown by \citetalias{Waters21}.  

Our rule of thumb from \S{1} that dynamical TI is stabilized in the acceleration
zone of outflows still leaves room for it to operate in highly bound regions that are likely dominated by magnetic pressure.  
A prime example is flow within vortices, as circulating streamlines aid the exponential growth 
of entropy modes by prolonging the time spent within a TI zone \citepalias[see][]{Waters21}.   
Unfortunately, we showed that for realistic UFO parameters, resolving 
the fastest entropy modes in 3D global simulations constitutes a problem requiring exascale computational resources (see \S{3.3}).  The scaling $N_l \propto \mathcal{M}_0^2$ implies that 1D simulations with an injection speed just twice as large (to make $v/c = 0.1$) carries four times the expense of our present calculations. 
It may therefore still be worth applying the local approximation first developed by \citet{PW15}.  

To afford global simulations, 
it will be important to develop a suitable refinement criterion to utilize
adaptive mesh refinement (AMR).  We note that simple criteria for detecting
density or temperature gradients are only of use for resolving clumps once they
are able to form.  Here the issue is that they cannot form unless the resolution
is high enough to resolve isobaric entropy modes advecting within the smooth background flow.  This requires a criterion 
designed to detect TI zones --- regions where $N_p < 0$ --- and the number of AMR levels needed is set by the resolution bound in \eqref{eq:Nlbound}.  While straight forward,
our uniform resolution study is necessarily a prerequisite for this development.

We have thus far failed to mention that our illustrative models do not have ionization parameters in the range $\log \xi = 3-6$, as needed to account for both the low and high ionization absorbers of multi-component UFO observations.  Reaching this regime using our current parameters is conceptually as simple as increasing $\Delta \Gamma/\Gamma_0$ from 2 to $10^3$, as $\xi$ will then increase by the same factor.  Because we took $\Gamma_0 = 10^{-3}$ here, this would correspond to near-Eddington rates.  However, the linear theory of TI is built around optically thin cooling functions, and the radiation field will no longer be everywhere optically thin in this regime (see \eqref{eq:Gamma_bound}).  Hence, this becomes a radiation hydrodynamics problem, where cloud opacities must be calculated to capture the dynamics properly \citep[e.g.,][]{Dyda20}.
Before undertaking this task, we plan to explore if the base of the flowtube can be placed on the `Compton branch' of the S-curve from the start (see Fig.~\ref{fig:A0}), as densities would be greatly reduced and the optically thin formalism would remain valid.  

We end our theory discussion by noting a few connections to the literature on dynamical TI in the circumgalactic medium (CGM) and intracluster medium (ICM).  
First, the importance of a ratio of dynamical and thermal timescales 
for assessing whether TI can operate was emphasized early on by
\citet{Efstathiou00} in the galactic wind environment, by
\citet{Pizzolato05} and \citet{Voit17} in analytic models of cooling flows, by
\citet{Binney09} in the context of high velocity clouds in the CGM, by
\citet{Gaspari12} and \citet{Li14} in simulations of AGN jets propagating into the ICM, and by
\citet{Sharma12} and \citet{McCourt12} in simulations of the CGM/ICM where the background state is a stratified hydrostatic atmosphere.
In moving from \eqref{eq:growth} to \eqref{eq:growth2}, we showed how such a ratio arises.  
We also derived a theoretical bound for this ratio (see \eqref{eq:tscale_bound}) that reveals under what circumstances multiphase gas production can occur even when $t_{\rm th} \gg t_{\rm dyn}$.

Second, the effects of $p\,dV$ work are likely the key reason 
for TI being more prevalent in cooling flows (or at least in the literature on these inflows) than in outflows: adiabatic heating in inflows tends to place the flow into the TI zone.  By contrast, adiabatic cooling makes it difficult for ionized AGN outflows to enter the TI zone, unless 
the solution offered in this work can be justified, which requires the local radiation field seen by the gas to be highly variable on timescales longer than $t_{\rm th}$.
We note also a further contrast with kiloparsec scale molecular outflows or supernova driven galactic winds \citep[see e.g.,][]{Veilleux20}, where gas must typically cool to enter a TI zone (because, respectively, the TI zone has $T < 10^4\,{\rm K}$ for neutral gas and the virialization temperature exceeds that of the TI zone for ionized gas).  Cooling by adiabatic expansion and heat advection are therefore conducive to clump formation in these large-scale outflows.  

Third, based on recent conclusions about the effects of cosmic ray (CR) physics on TI in the CGM/ICM \citep[e.g.][]{KQ20,Butsky20,Tsung21}, it will be important to determine the circumstances under which CR heating or pressure can alter the outcome of our simulations.  According to \citet{Wagner05}, the isobaric stability criterion for TI is unchanged, irrespective of the CR diffusivity or pressure.   
This means that the boundary of the TI zone on the phase diagram in Fig.~\ref{fig:A0} (i.e. the Balbus contour) will not change either.  The actual growth rates can be affected, however.  As shown by \citet{KQ20}, the growth rate of isobaric entropy modes decrease significantly with increasing CR pressure, while it is quite insensitive to the CR heating rate.  More interestingly, they found that unless the CR transport is highly diffusive, an entropy mode ceases to be simply advected with the flow; it begins to propagate as an overstable mode at a fraction of the Alfven speed, that fraction being proportional to the CR to gas pressure ratio.  This would tend to suppress clump formation in highly supersonic flows, as the entropy modes would spend even less time occupying a TI zone.  

\subsection{Observational implications}
Neither of the two commonly referenced scenarios that can explain the presence soft X-ray absorption in UFO observations, namely post-shock cooling models \citep{King10,Pounds13,FGQ12} and cloud entrainment models \citep[e.g.,][]{Serafinelli19}, can account for the very recent discovery of instances in which the soft and hard X-ray absorbers have the same blueshift \citep{Reeves18,Reeves20}.  Decades of shock-cloud interaction studies have revealed how difficult it is to accelerate cold gas to highly supersonic speeds \citep[e.g.,][and references therein]{BB20}, while \citet{Wagner13} performed hydrodynamical simulations of a $0.1c$ UFO plowing into a multiphase medium and demonstrated that the entrained clouds reached speeds no greater than $300\,\rm{km\,s^{-1}}$.
The entrained ISM scenario is further ruled out by the short-term variability in the column density measurements of the low ionization component in these recent observations.

Here we showed that dynamical TI naturally leads to clumps moving at the same speed as the wind.
It is crucial to note that the VW solution displays strong ionization stratification in the background flow solution, with $\xi$ decreasing by over three orders of magnitude.  Therefore, the VW clumps are not needed to account for low ionization gas at UFO speeds.  Rather, they are responsible for the high ionization component, as gas with $\log\xi > 3$ is produced as a result of the runaway heating accompanying TI.  
The column density of the $\log\xi > 3$ component is small, but that could change in multi-dimensional calculations.  

While we modeled the increase in flux necessary to place gas in a TI zone as a rise in luminosity,
this rise can be completely unrelated to the X-ray light curve.  For example, if the observed X-ray luminosity is relatively constant at a $\Gamma$ corresponding to that after applying the $\Delta \Gamma$, our solutions are simply demonstrating the effects of an \textit{unobserved} deshadowing event.  
This is the reason our sketch in Fig.~\ref{fig:sketch} includes a penumbra; the dotted line is envisioned to sweep out very different angles with time.  
One such scenario that would cause this is commonly invoked in the literature: the formation of a layer of `self-shielding' gas that may be a recurring transient. 

The role of shadowing can be important in regards to blueshift variability also.  Very recently, \citet{Braito21} reported on the discovery of unprecedented velocity variations within the Fe K band: over a period of just 16 days, the measured blueshift decreased by nearly a factor of 3 from $v/c \approx 0.2$ to $v/c \approx 0.07$.  
They attributed this to the disk wind having a non-axisymmetric velocity field, which would require the absorber to be located within $10^3\,r_g$.
Shadowing can lead to strong blueshift variability in lines produced much farther out, as a result of partial obscuration of the continuum source.  Taking the geometry in Fig.~\ref{fig:sketch} as an example, a drop in velocity can occur when the penumbra extends past the RW flowtube, so only the VW flowtube, with its smaller line of sight velocity, remains observable.  This scenario entails that the line equivalent width and the continuum flux be positively correlated.  Note the clumps play no role here, other than to provide a soft X-ray absorption component if necessary.  

The observational implications of relying on a non-constant radiation field to place gas into a TI zone are equally noteworthy: ionization changes due to increases in the ionizing flux \emph{should lead to} the production of clumps, and hence to partial covering that will cause further absorption line variability.  In other words, what is often taken in the literature to be competing interpretations of the spectral variability --- the gas ionization responding to changes in the radiation field versus denser clumps evolving along or crossing the line of sight --- should in fact be viewed as tightly coupled processes.  

\section{Summary and conclusions}
In this work, we revisited the theory of dynamical TI, first developed in the context
of cooling flows in galaxy clusters, and extended it to ionized AGN outflows.
Here we summarize our main results.
\begin{itemize}[leftmargin=*,topsep=0pt,itemsep=-1ex,partopsep=1ex,parsep=1ex]

\item
The most favorable conditions for dynamical TI are found at radii beyond an outflow's acceleration zone. 
In contrast, our recent thermal wind studies showed that the parameter space for dynamical TI to operate is quite narrow because thermal driving is accompanied by strong acceleration.  Velocity gradients can stabilize TI by stretching individual entropy modes, which inhibits exponential growth.   

\item Even mildly relativistic outflows can become clumpy.  
For highly supersonic flows that have reached a terminal velocity (with energetics characterized by $\rho v^2 \gg |\Phi_{\rm eff}|$ as a result of radiation or magnetic launching as well as $\rho v^2 \gg p$ so that thermal driving is negligible), entropy modes are not subject to disruption, their growth being limited by the time spent in a TI zone but otherwise set by the linear theory rate for a static plasma. 
We showed using the ionization parameter framework that the resolution necessary to capture the fastest growing entropy modes in hydrodynamical calcuations increases as the Mach number \textit{squared}.
For the parameter regime explored here, performing even 2D AMR-based simulations will require computational resources similar to that needed in current 3D global AGN disk simulations.

\item Time variability in the radiation field is a robust means of placing gas in a TI zone.  In the paradigm of a `cold', thin, optically thick AGN disk, gas in the disk atmosphere is expected to occupy the cold branch of the S-curve at temperatures $T \lesssim 10^5\,{\rm K}$.  If this atmosphere is the base of an outflow, the plasma can enter a TI zone because $\xi$ increases in proportion to the flow velocity (see \eqref{eq:FvA}).  However, entering a TI zone as a result of acceleration is not conducive to clump formation (as discussed in the first bullet point above).  In the absence of velocity gradients, $\xi$ tends to decrease due to adiabatic cooling or heat advection.  This tendency simply persists in the presence of temporal increases in the radiation flux, allowing $\xi$ to increase overall.
In our models, a factor of 3 rise in the flux was sufficient to place the wind fully into a TI zone, whereas increases by a factor of 50 have been observed in UFOs \citep[e.g.,][]{Parker17}.

\item Clump formation from dynamical TI is permitted for $t_{\rm th} \gg t_{\rm dyn}$.
At UFO speeds, the ratio $t_{\rm dyn}/t_{\rm th}$ can be less than one in much of the flow (see \eqref{eq:trat}), as was found for the RW case here.  This is in agreement with the studies on dynamical TI in the CGM and ICM mentioned in \S{4}, where it was first shown that multiphase gas production is not prohibited even for $t_{\rm dyn}/t_{\rm th} \sim 0.1$.  However, this has been an empirical result, whereas in \S{2} we derived a theoretical lower bound on $t_{\rm dyn}/t_{\rm th}$ (see \eqref{eq:tscale_bound}).
Consistent with our analysis accompanying this lower limit, recent studies have cautioned against elevating the value of $t_{\rm dyn}/t_{\rm th}$ to that of a robust diagnostic of when dynamical TI can saturate \citep[e.g.,][]{Choudhury19, Esmerian21}.
We emphasize that knowing the value of this ratio is no substitute for evaluating the actual criterion for gas to be thermally unstable, which is that it occupy a TI zone (see \S{2}). 

\item Isobaric perturbations can give rise to non-isobaric clouds.
By design, the RW and VW flowtubes have identical conditions at the base,
so any differences in the saturation of dynamical TI results from the properties of the background flow solutions. By focusing on the evolution of a single, short wavelength perturbation inserted periodically into the flow, we identified, in the case of VW clumps, the telltale signs of non-isobaric behavior.  In the local TI approximation, this behavior is unique to large wavelength perturbations \citep{Waters19a}.  
Non-isobaric dynamics persisted in the VW flowtube, even once the clumps occupied the cold branch of the S-curve and were no longer visible in the temperature profile.  
\end{itemize}

We conclude by re-framing the first two bullet points above to emphasize an elegant and remarkably simple theoretical connection between the warm absorber and UFO parameter regimes according to the theory of dynamical TI.  By the special case of Crocco's theorem that holds along streamlines, $\nabla Be = T\nabla s$, there are just two ways to have $\nabla s = 0$, the condition most conducive to multiphase gas production (because steep entropy gradients are associated with the disruption of unstable entropy modes; see \S{4}): $Be \approx constant$ or $Be = 0$.  
The latter case was explored by \citetalias{Dannen20} and \citetalias{Waters21} and corresponds to the distances and speeds of warm absorbers.  The case $Be \approx constant$ was shown here to apply to regions beyond the acceleration zone of UFOs.

\acknowledgements 
We dedicate this paper to the memory of Bill Mathews (March 21, 1937 - September 24, 2021), who, in addition to making pioneering contributions to the study of AGN clouds (see \S{1}), published the amplification solution that provided a basis for this study (see \eqref{eq:growth}).
We thank James Reeves and George Chartas for email correspondence on the latest findings with regards to UFO observations, Tim Kallman for encouraging us to closely examine what causes an irradiated flow to depart from the S-curve, and the anonymous referee for a constructive report and for pointing us to relevant work on the effects of cosmic rays.
Support for this work was provided by the National Aeronautics and Space Administration under ATP grant NNX14AK44G.

\appendix 
\section{Constraints on the governing parameters}
The wind base is assumed to occupy the cold branch of the S-curve, so this constraint is
\beq \Xi_0 < \Xi_{\rm c,max}. \seq  
For our S-curve, $\Xi_{\rm c,max} \approx 10^{1.1}$.
Our modeling framework requires the flow to have an optically thin line of sight to the X-ray source.  At minimum, this requires $\tau_0 = n_0 r_0 \sigma_{\rm es} < 1$, and note that $\tau_0$ is the actual scattering optical depth only if $n_0$ is constant within $r_0$.  With $n_0 = f_X \Gamma_0 \bar{m} c^2 r_g/(k T_0 r_0^2 \sigma_{\rm es} \Xi_0)$ by \eqref{eq:Xi0}, the bound $n_0 < (r_0 \sigma_{\rm es})^{-1}$ gives
\beq
\Gamma_0 < 1.53 \times 10^{-3} f_X^{-1}\left(\f{T_0}{10^5\, {\rm K}}\right)^{-1} \left(\f{\Xi_0}{10}\right) \left(\f{r_0/r_g}{10^4}\right) \label{eq:Gamma_bound}. \seq
Considering that UFOs are likely accelerated within $10^4\,r_g$,
it follows from \eqref{eq:HEP0} that $r_0/r_g < 10^4$ corresponds to
\beq {\rm HEP}_0 >  3.92\times 10^3 \left(\f{T_0}{10^5\, {\rm K}}\right)^{-1}(1 - \Gamma) \label{eq:HEP_bound}. \seq
It remains to place a bound on $\mathcal{M}_0$.
By Bernoulli's theorem in a non-adiabatic regime, 
\beq \gv{v}\cdot \nabla Be = -\mathcal{L} + \gv{v}\cdot \gv{f}_b , \seq
we see that the Bernoulli function $Be = v^2/2 + c_s^2/(\gamma - 1) + \Phi_{\rm eff}$ is defined independently of an external body force, $\gv{f}_b$, here taken to be the sum of the Lorentz force and the radiation force due to line-driving that accelerated the flow.  A bound on $\mathcal{M}_0$ follows from the energetic requirement that the flow be unbound beyond the acceleration zone where $\gv{f}_b$ is negligible, i.e. from $Be > 0$.
Dividing $Be$ by $|\Phi_{\rm eff}|$ and evaluating it at $r=r_0$ casts $Be_0 \equiv Be(r=r_0)$ in terms of the governing parameters: 
\beq
\f{Be_0}{|\Phi_{\rm eff}(r_0)|} = \f{1}{2}\f{\mathcal{M}_0^2}{{\rm HEP}_0} + \f{1}{\gamma - 1}\f{1}{{\rm HEP}_0} - 1.
\label{eq:Be}
\seq
The middle term above is negligible because ${\rm HEP}_0 \gg 1$ by \eqref{eq:HEP_bound}.  Hence, ${Be_0 > 0}$ is equivalent to
\beq 
\mathcal{M}_0 > \sqrt{2\,{\rm HEP}_0} \gtrsim 88.5 \sqrt{1-\Gamma} \left(\f{T_0}{10^5\, {\rm K}}\right)^{-1/2},
\label{eq:M0bound}
\seq
i.e. the outflow must be highly supersonic.


\end{document}